\documentclass[12pt]{iopart}
\usepackage{iopams}
\usepackage[latin1]{inputenc}
\usepackage[T1]{fontenc}
\usepackage{ifpdf}
\ifpdf
\usepackage[pdftex]{graphicx}
\else
\usepackage[dvips]{graphicx}
\fi
\usepackage{blackdvi}
\usepackage{cite}
\usepackage{lineno}
\ifpdf
\usepackage[pdftex]{hyperref}
\else
\usepackage[dvips]{hyperref}
\fi

\def \Kla#1{\left( #1 \right)}
\def \Klb#1{\left[ #1 \right]}
\def \KlB#1{\left| #1 \right|}

\def \Unit#1{\thinspace\hbox{\rm #1}}
\def \SCIe#1#2{\hbox{$#1\times 10^{#2}$}}

\def \varSB{\mathop{\rm var}\nolimits}

\def \Ee{E_{\rm e}}

\begin{document}
\title{New precision measurements of free neutron beta decay with cold neutrons}
\author{S~Bae\ss{}ler$^{1,2}$, J~D~Bowman$^2$, S~Penttil\"a$^2$, D~Po\v{c}ani\'c$^1$}
\address{$^1$ University of Virginia, Charlottesville, VA 22904, U.S.A.}
\address{$^2$ Oak Ridge National Laboratory, Oak Ridge, TN 37831, U.S.A}
\ead{baessler@virginia.edu}

\begin{abstract}
Precision measurements in free neutron beta decay serve to determine the coupling constants of beta decay, and offer several stringent tests of the Standard Model. This paper describes the free neutron beta decay program planned for the Fundamental Physics Beamline at the Spallation Neutron Source at Oak Ridge National Laboratory, and puts it into the context of other recent and planned measurements of neutron beta decay observables.
\end{abstract}
\pacs{23.40.Bw,13.30.Ce,12.15.Hh}
\submitto{\jpg}
\maketitle

\section{Introduction}
Measurement of observables in free neutron beta decay falls within the broader field of study of the fundamental properties and symmetries of the electroweak interaction at low energies. Although successful without parallel, the present Standard Model (SM) of elementary particles and their interactions, based on the $SU(3)_{\rm C}\times SU(2)_{\rm L}\times U(1)_{\rm Y}$ gauge symmetries, is known to be incomplete. Additional particles and phenomena must exist. Questions regarding possible extensions of the SM are being simultaneously addressed at the high energy frontier, using particle colliders, and at the precision frontier at low and intermediate energies. Precision measurements in neutron beta decay belong in the latter category, and seek to illuminate questions related to the number of quark generations (through exploring quark-lepton universality via the Cabibbo-Kobayashi-Maskawa, or CKM matrix), non-SM forms of weak interaction (i.e., scalar, pseudoscalar and tensor), as well as trace evidence of supersymmetry.
The redundancy inherent in the SM description of the neutron beta decay process allows uniquely sensitive checks of the model's validity and limits \cite{Dub91, Byr95, Pro07, Kon10, Bhat12, Cir13, Hol14}, with strong implications in astrophysics \cite{Dub11}. 

In this work we discuss active and recent free neutron beta decay experiments. We restrict ourselves to experiments that have either obtained results in the last decade, or are poised to do so, and which use cold neutrons. Neutrons are considered cold if they are extracted in a beam from a
     cold source in which their kinetic energy is in approximate
     equilibrium with matter at a low temperature (usually about $40$\Unit{K});
the neutron's average kinetic energy is about $5$\Unit{meV}. Our main emphasis is on the description of the experiments that are under construction for the Fundamental Physics Beamline (FNPB) of the Spallation Neutron Source (SNS) at Oak Ridge National Laboratory. This paper is part of a focus issue on precision semileptonic weak interaction physics. The theoretical motivation and relevance of the various experiments to contemporary particle / nuclear theory is summarized in \cite{Hol14} in this issue. Neutron beta decay experiments that use ultracold neutrons (neutrons with an average energy about $100$\Unit{neV} that can be stored in material or magnetic bottles) are discussed in \cite{Young14} in this issue. Experiments that probe time-reversal ($T$) invariance are discussed in \cite{Sev14} in this issue, along with beta decay experiments involving nuclei. Recent reviews of neutron beta decay that include more of the history have been given in \cite{Sev06,Abe08,Nico09}. 

     Most past and proposed neutron beta decay experiments focus on
     determining one or more of the correlation coefficients defined by
     Jackson \etal \cite{Jack57}, who parametrized the differential
     decay rate to leading order as 
\begin{equation}   
\fl  ~\qquad
\frac{d^3\Gamma}{d\Ee d\Omega_{\rm e} d\Omega_\nu}  \propto 
    p_{\rm e} E_{\rm e} \Kla{E_0-E_{\rm e}}^2
\xi \cdot \left( 1+a\frac{{\vec p_{\rm e}}\cdot{\vec p_\nu}}{\Ee E_\nu}+b\frac{m_{\rm e}}{\Ee} \right. 
 \left. +{\vec{\sigma}}_{\rm n}\cdot\Klb{A\frac{\vec p_{\rm e}}{E_{\rm e}}+B\frac{{\vec p_\nu}}{E_\nu}} \right),
\label{eq:CorrCoeffDefinition}
\end{equation}   
where $p_{\rm e}, p_\nu, E_{\rm e}, E_\nu$, are the relativistic momenta and energies of decay electron and neutrino, and $E_0$ is the endpoint energy of the electron. The quantity $\sigma_{\rm n}$ denotes the neutron spin. 
In the SM, we have $\xi = G_{\rm F}^2V_{\rm ud}^2 (1+3\lambda^2)$. The Fermi constant, $G_{\rm F}$, best measured in muon decay, was
   recently determined to be $G_{\rm
   F}=\SCIe{1.1663787(6)}{-5}$\Unit{GeV$^{-2}$} by the MuLan
   collaboration \cite{MuLan13}.
$V_{\rm ud}$ is the relevant element of the Cabbibo-Kobayashi-Maskawa (CKM) matrix. The ratio of the axial vector ($g_{\rm A} = G_{\rm F}V_{\rm ud}\lambda$) and vector ($g_{\rm V} = G_{\rm F}V_{\rm ud}$) coupling constants, $\lambda= g_{\rm A}/g_{\rm V}$ takes into account the renormalization of the axial vector current by the structure of the nucleon. 
The calculation of $\lambda$ from first principles is too uncertain, and is therefore usually treated as a free parameter in the SM.
   Key observables are defined through \eref{eq:CorrCoeffDefinition}, namely $a$, the neutrino-electron
   correlation coefficient, $b$, the Fierz interference term, $A$, the beta asymmetry,
   and $B$, the neutrino asymmetry.
    Additional terms appear if outgoing particle spins are detected, or
    if time-reversal ($T$) invariance is not conserved. In the low-energy limit of the SM, neutron beta decay is described as a $V-A$ interaction, requiring the Fierz interference term $b$ to vanish, and
giving the coefficients $a$, $A$, and $B$ non-zero values that depend on a single parameter, the value of $\lambda$ (here assumed to be real in the absence of $T$ violation). These dependencies are detailed in \tref{tab:SMValues_CorrCoeff}. The $a$ and $A$ coefficients are most sensitive to $\lambda$.

\begin{table}[!ht]
  \caption{Observables (Obs.) in neutron beta decay, their Standard Model value (in lowest order), results from direct measurements, and sensitivities to $\lambda$. $f^{\rm R}$ is a phase space factor that includes part of the radiative corrections, and $m_{\rm e}$ is the electron mass.}
  \label{tab:SMValues_CorrCoeff}
\begin{center}
\begin{tabular}[c]{ccccc} 
\br
Obs. & SM prediction & Dir.\ meas't. \cite{PDG12} & Sensitivity to $\lambda$\\ 
 \mr \\[-10pt]
$\tau_{\rm n}$ 
   & $\displaystyle \frac{2\pi^3 \hbar^7}{f^{\rm R}m_{\rm e}^5 
        c^4 G_{\rm F}^2 V_{\rm ud}^2\Kla{1+3\lambda^2}}$ & $880.0(9)$\Unit{s} 
                                                               & --- \\[12pt]
$a$ 
   & $\displaystyle \frac{1-\lambda^2}{1+3\lambda^2}$ 
       & $-0.1030(40)$ 
           & $\displaystyle\frac{{\rm d}a}{\rm d\lambda} \sim -0.30$ \\[9pt]
$b$ 
   & $0$ 
       & (see \sref{sec:FierzTerm}) 
           & $\displaystyle\frac{{\rm d}b}{\rm d\lambda} = 0$ \\[9pt]
$A$ 
   & $\displaystyle -2\frac{\lambda^2+\lambda}{1+3\lambda^2}$ 
       & $-0.1176(11)$ 
           & $\displaystyle\frac{{\rm d}A}{\rm d\lambda} = -0.37$ \\[12pt]
$B$ 
   & $\displaystyle 2\frac{\lambda^2-\lambda}{1+3\lambda^2}$ 
       & $0.9807(30)$ 
           & $\displaystyle\frac{{\rm d}B}{\rm d\lambda} = -0.076$ \\[6pt]
\br
\end{tabular}
\end{center}
\end{table}
 
\section{The neutron lifetime}

Precision measurements of at least one neutron decay correlation coefficient and of the neutron lifetime are needed to determine SM parameters with neutron decay, $V_{\rm ud}$ and $\lambda$. Furthermore, the value of the neutron lifetime is a critical input for the computation of nucleon abundances in primordial nucleosynthesis \cite{PDG12_BBN}. We discuss the experiments that measure the neutron lifetime first. Neutron lifetime measurements use one of two methods, each with significantly different systematic errors. The first approach requires counting neutron decays in a fiducial volume containing neutrons (usually from a
     cold neutron beam), and comparing the result with the number of neutrons
     present in that same volume (the ``beam method''). Alternatively, one stores ultracold neutrons in a bottle, and observes the decay of the number of neutrons with time (the ``storage method''). In the beam method, the neutron lifetime is measured as $\tau_n=N/r$, where $N$ is the number of neutrons present in the fiducial volume
     at any time, and $r$ is the rate of the detected decays. 
In the storage method, the number of neutrons in a bottle decreases with time $t$ through $N(t)=N_0 \exp\Kla{-t/\tau_{\rm n, storage}}$.
For the storage method to work, the efficiency of the neutron detection does not have to be precisely known. The main issue in the storage method is to estimate the loss rate of neutrons through processes other than beta decay, e.g., through capture or upscattering when bouncing off the bottle wall.  This is complicated by the fact that the loss mechanisms are not all understood from first principles; see \cite{Poko99,Lamo02,Nes13} for recent discussions. The rate of wall losses appears not to be constant over the neutron phase space, which causes the phase space density to evolve with time, which, in turn, causes the effective wall loss rate and/or detection efficiency of the neutrons to change with time. Only ultracold neutrons, can be stored in a bottle.  This paper discusses experiments with cold neutrons; we direct the reader to Young \etal \cite{Young14} for a discussion of neutron lifetime experiments that use the ``storage method'', or to Wietfeldt \etal \cite{Wietf11}.

\begin{figure}[!ht]
\begin{center} \includegraphics{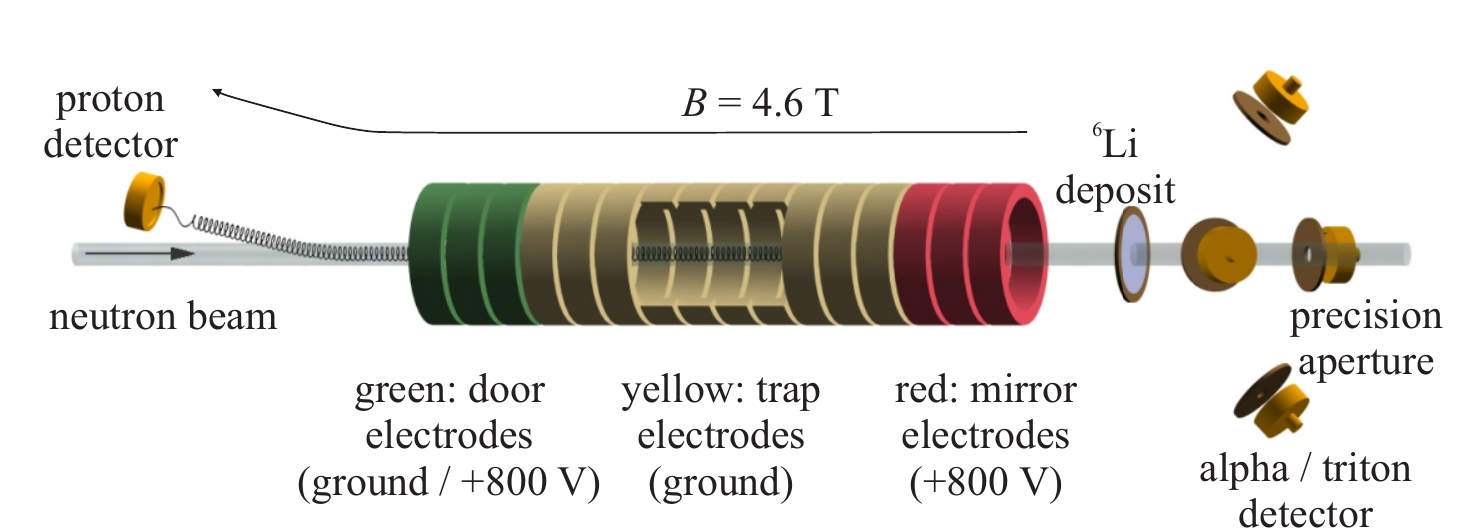} \end{center}
\caption{Setup of the most recent neutron lifetime experiment at the cold neutron beamline NG-6 at NCNR \cite{Nico05}.}
\label{fig:NIST_taun_scheme} 
\end{figure}

Recent measurements of the neutron lifetime with the beam method have been performed at the NG-6 beamline at the National Institute for Standard and Technology (NIST) Center of Neutron Research (NCNR) \cite{Nico05}. \Fref{fig:NIST_taun_scheme} depicts a recent NCNR apparatus, similar to the one built earlier by Byrne \etal~\cite{Byr90,Byr96}. In the NCNR experiment, a cold neutron beam is directed through a strong axial magnetic field. A stack of cylindrical electrodes, aligned axially, is used to create an electrostatic potential that is uniform in a central region, and has a positive voltage higher than required by the proton endpoint at its ends, labeled as ``mirror electrodes'' and ``door electrodes'' in the figure. The central region is the fiducial volume where a number of neutron beam particles decay in flight. Protons from these neutron decays are trapped, radially by the magnetic fields, and axially by the electric field; the electromagnetic field setup is essentially a cylindrical Penning trap. The voltage of the door electrodes is lowered after collecting decay protons in the trap for a preset time. Decay protons escape through the door, and are guided to a silicon surface barrier ``proton'' detector on a negative high voltage where they are counted.  A set of measurements with different trap lengths is performed in order to avoid the problem of imprecisely known trapping efficiency for decay protons created close to one of the electrostatic barriers. The variation of the proton count rate with the trap length is used to determine the number of neutron decays per unit length of the neutron beam.

The challenge for experiments that use the beam method is to combine the absolute measurement of the number of protons detected with an absolute measurement of the number of neutrons in the fiducial volume, or, equivalently, a measurement of neutron beam capture flux, that is, the neutron beam flux weighted with $v_0/v$, where $v$ is the neutron velocity and $v_0=2200$\Unit{m/s} the average velocity of a thermal neutron.  Both the probability for a beam neutron to decay within the fiducial volume, and the probability for neutron capture in a thin target, follow a $1/v$ law. In \cite{Nico05}, the neutron beam capture flux was measured with a neutron monitor made from a well-characterized $^6$Li deposit. The detectors surrounding the deposit counted the rate of alpha particles and tritons produced in the $^6$Li(n,$\alpha$)$^3$H reaction. The main uncertainty in the experiment was the knowledge of the cross section for this reaction, which is $\sigma= 941.0(13) \Unit{b} \cdot v_0/v$ \cite{ENDF93}. 
The $^6$Li deposit neutron monitor efficiency was calibrated using a monoenergetic beam in a later study. The neutron monitor count rate was compared to the rate in an absolute counter for thermal and cold neutron beams, as described in \cite{Gill89}. The absolute counter relies on an alpha-gamma spectrometer directed to a totally absorbing $^{10}$B deposit in the beam.  Neglecting small corrections, all neutrons in the beam react with the deposit, and produce an alpha and one or more gammas that are detected in the counters.  The alpha and gamma detection efficiencies can be found by comparing coincidence and single rates in the different counters.  Using this absolute calibration, and after taking into account the neutron beam profile and neutron energy distribution, a small correction to the previous neutron lifetime result of \cite{Nico05}, $+1.4(5)$\Unit{s}, has been made in \cite{Yue13}. 

\begin{figure}[!htb]
\begin{center} \includegraphics{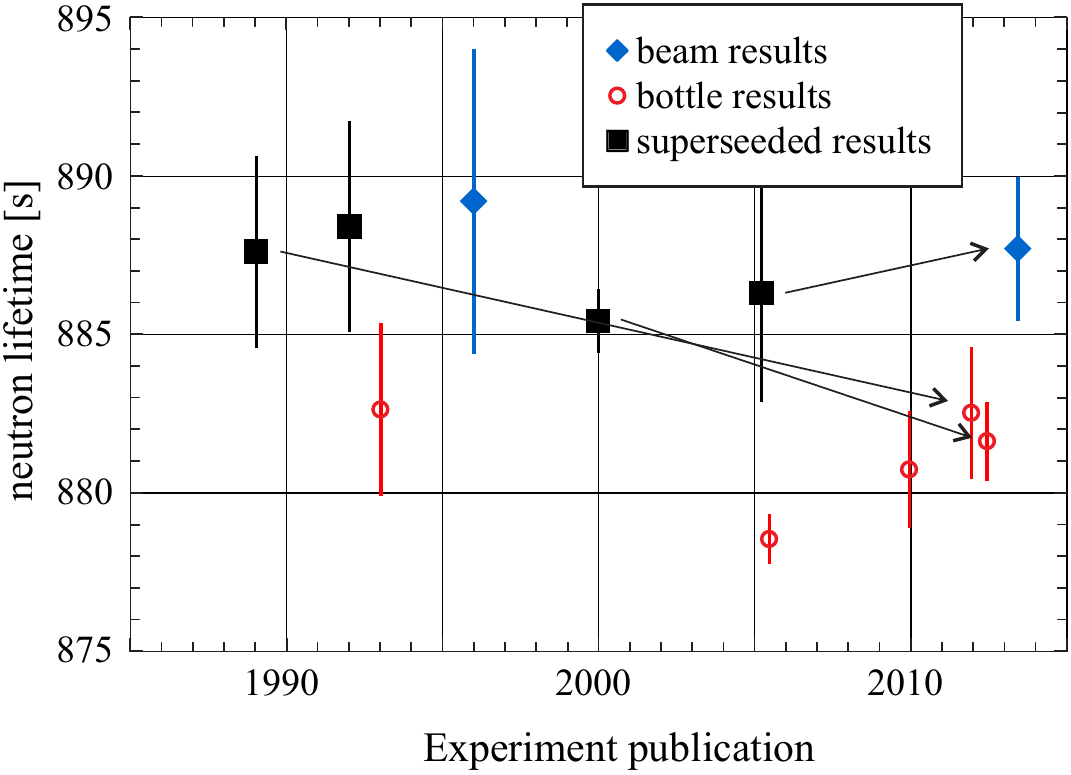} \end{center}
\caption{Compilation of the most recent measurements of the neutron lifetime. Data from two beam experiments are shown as blue diamonds, from Byrne \etal ~\cite{Byr96}, and Yue \etal ~\cite{Yue13} (from older to newer). Data from five bottle experiments are shown as red circles, from Mampe \etal ~\cite{Mam93}, Serebrov \etal ~\cite{Ser05}, Pichlmaier \etal ~\cite{Pichl10}, Steyerl \etal ~\cite{Stey12}, and Arzumanov \etal ~\cite{Arz12}. Previous results that have been retracted, re-analyzed, or otherwise improved are shown as black squares, from Mampe \etal ~\cite{Mam89}, Nesvizhevsky \etal ~\cite{Nes92}, Nico \etal ~\cite{Nico05}, and Arzumanov \etal ~\cite{Arz00}.}
\label{fig:nLifetimeOverview} 
\end{figure}

\Fref{fig:nLifetimeOverview} summarizes current experimental data on the neutron lifetime. There is a discrepancy of several seconds between the average results from neutron lifetime measurements using the two different methods, which points to an unknown systematic uncertainty in at least one of the techniques. Attempts have been made for more than a decade to add precise measurements using magnetically stored ultracold neutrons in the expectation that in those, the losses in wall reflections are negligibly small. Ezhov \etal \cite{Ezh05} and Salvat \etal \cite{Salv14} have recently achieved magnetic storage. Both experiments cannot resolve the discrepancy (yet). Salvat \etal have the goal to reach a precision of one second soon. The next generation of beam lifetime measurements at NCNR aims to reach the same level. Wietfeldt \etal~\cite{Wiet14} argue that the beam method has the potential to be substantially improved even beyond that.

\section{The neutrino-electron correlation {\normalfont \bfseries \itshape a}}

Several different strategies have been brought forward to determine the $a$ coefficient. The most straightforward way would involve detection of the neutrino, which is not feasible. One possibility is to infer the $a$ coefficient from the proton spectrum. A different method is the use of a combination of information on proton and electron momentum to extract the angle between neutrino and electron. 


\subsection{The neutrino-electron correlation {\normalfont \bfseries \itshape a} from the proton spectrum}

\begin{figure}[!ht]%
   \begin{center} \includegraphics{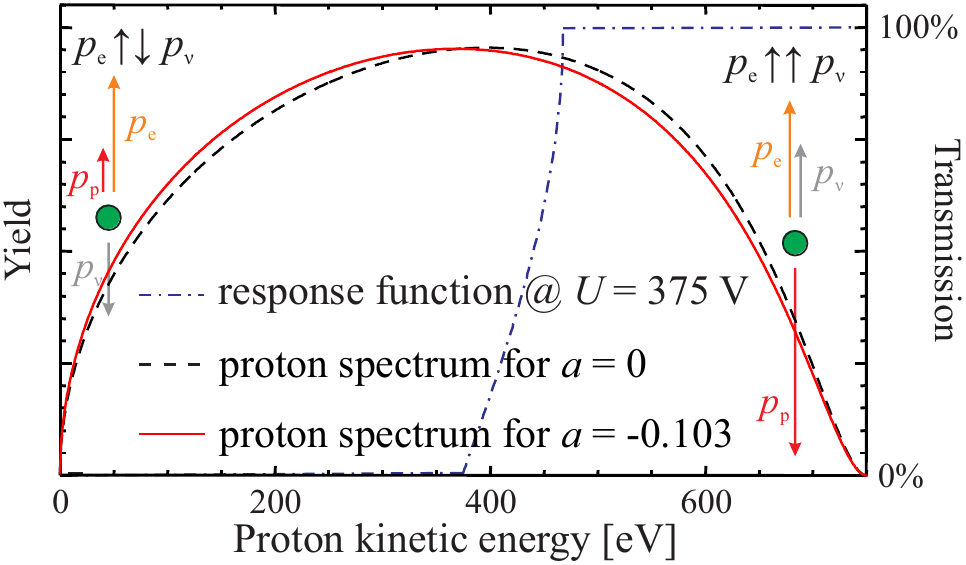} \end{center}
   \caption{Proton spectrum in free neutron beta decay for the experimentally observed value of the $a$ coefficient (red solid line), and for a vanishing $a$ coefficient (dashed black line). The blue dashed dotted curve is plotted in reference to the right axis, and is explained in the discussion of the $a$SPECT experiment.}  
   \label{fig:ProtonSpectrum}
\end{figure}

The shape of the spectrum follows from the three-body decay kinematic of the decay of the free neutron at rest. A positive value for the $a$ coefficient would favor alignment of the electron and neutrino momenta. Nature chose a small negative value $a=-0.1030(40)$ \cite{PDG12}. In order to balance the momenta of the decay particles,  protons at the high energy end of the proton spectrum are less abundant than they would be for a large positive value for the $a$ coefficient. Proton spectra for two different values of $a$ are shown in \fref{fig:ProtonSpectrum}.

The most precise results for the $a$ coefficient are extracted from measurements of the proton spectrum. Stratowa \etal obtained $a=-0.1017(51)$ by detecting the energies of protons created in neutron decays occurring in a vacuum tube that passes close to the core of a nuclear reactor \cite{Str78}.

\begin{figure}[!ht]%
   \begin{center} \includegraphics{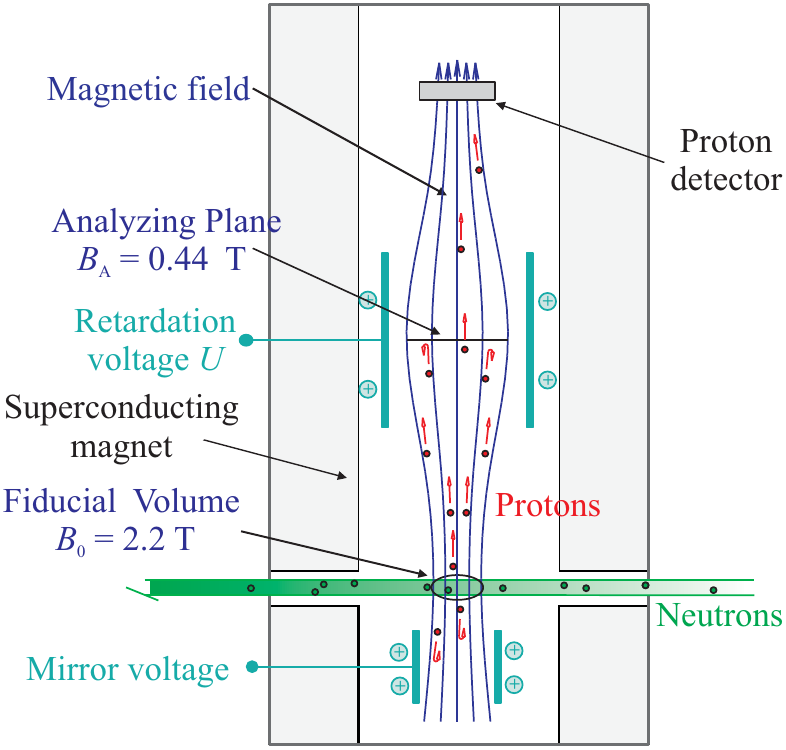} \end{center} 
   \caption{Schematic diagram of the $a$SPECT spectrometer. Details are explained in the text. } 
   \label{fig:aSPECTSketch}
\end{figure}

Byrne \etal and the newer experiment $a$SPECT use variants of a retardation spectrometer. The former experiment is finished, and reported $a=-0.1054(55)$\cite{Byr02}. The $a$SPECT experiment was proposed in \cite{Zim00} and its design was further explained in \cite{Glu05}. The configuration and the principle of operation of the $a$SPECT spectrometer are illustrated in  \fref{fig:aSPECTSketch}. Unpolarized cold neutrons are directed through the sensitive volume of a magnetic spectrometer. About one of $10^8$ neutrons decays in the fiducial volume. The recoil protons from the decay are gyrating around the magnetic field lines. Protons emitted downward are reflected by an electrostatic mirror, so all decay protons move upward at some point, towards the analyzing plane and the proton detector. A variable retardation voltage is applied at the analyzing plane. Only protons with sufficient kinetic energy can pass, and those are accelerated by a high voltage of about -15\Unit{kV} and magnetically focused onto a proton detector that counts the number of arriving protons. 

\begin{figure}[!ht]%
  \begin{center} \includegraphics{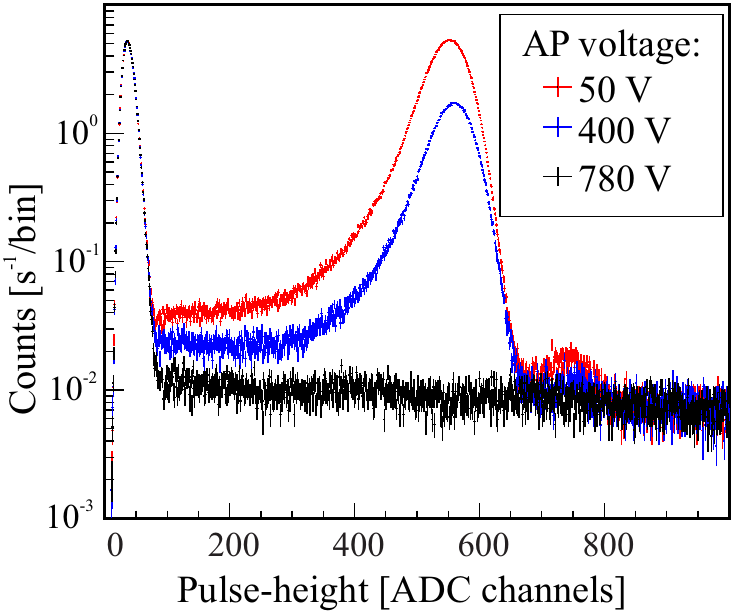} \end{center}
   \caption{Proton spectra in the proton detector for different retardation voltages $U$ in the analyzing plane, after \cite{Sim10}.}  
   \label{fig:SDDProtonSpectra}
 \end{figure}

A silicon drift detector (SDD) is used to detect the protons \cite{Gatti84}. An SDD is a semiconductor
detector based on the principle of sidewards depletion which allows the full depletion of a
large detector volume with a very small readout node. The smallness of the readout electrode leads to an improved noise performance compared to a PIN diode with the same dimensions. The price to pay is the long drift path, mostly sideways, of the secondary electrons, with drift times of the order of microseconds. This can be used to achieve position resolution if the time of impact of the incoming particle is known, but otherwise limits the accuracy in the determination of the impact time.
The SDDs used in the aSPECT spectrometer have been manufactured by PNSensor GmbH \cite{PNSensor}. One SDD chip contains a row of three square pads with a side length of 1\Unit{cm}. Each pad has a thickness of 450\Unit{$\mu$m}, and is covered by a aluminum layer of 30\Unit{nm} thickness for protection. To reduce the electronic noise even further, the first FET is integrated on the SDD chip. In \fref {fig:SDDProtonSpectra}, pulse height spectra from the proton detector for different retardation voltages are shown. The signal from incoming protons with about 15\Unit{keV} is well separated from the noise. Integration over the proton peak gives the proton count rate, which has been taken for several analyzing plane voltages to determine the proton spectrum shape. 
A measurement with a voltage above the proton endpoint energy (about 750\Unit{eV}) can be used to subtract the background. \Fref{fig:ProtonSpectrum} indicates how a measurement of the proton spectrum shape translates into a measurement of the correlation coefficient $a$.

The analysis has to take into account the gyration of the protons, that is, the fact that the proton orbit is a combination of circular motion around the field lines and axial motion along them. The protons experience a decreasing magnetic field on the path towards the analyzing plane, which causes their momenta to align with the magnetic field lines. The transmission function describes the probability for protons to traverse the electrostatic potential in the analyzing plane; they pass if their energy in the axial motion is above $eU$. For a typical setting of the retardation voltage, the transmission function vanishes for proton energies below $eU$. It reaches 100\% at high proton energies, where we have sufficient kinetic energy in the axial motion to pass the analyzing plane. In an intermediate range of proton energies, only a fraction of protons can pass, i.e., the ones whose momenta are most aligned with the magnetic field. The transmission function can be analytically calculated in the adiabatic approximation; and the input parameters needed are the electrostatic potential and magnetic field values in both fiducial volume and analyzing plane. As an example, the transmission function for $U=375$\Unit{V} is shown as the dashed dotted line in \fref{fig:ProtonSpectrum}. A detailed field map is only needed to ensure that the field is sufficiently smooth for the adiabatic approximation to be valid.
The analysis has to take into account the gyration of the protons, that is, the fact that the proton orbit is a combination of circular motion around the field lines and axial motion along them. The protons experience a decreasing magnetic field on the path towards the analyzing plane, which causes their momenta to align with the magnetic field lines. The transmission function describes the probability for protons to traverse the electrostatic potential in the analyzing plane; they pass if their energy in the axial motion is above $eU$. For a typical setting of the retardation voltage, the transmission function vanishes for proton energies below $eU$. It reaches 100\% at high proton energies, where we have sufficient kinetic energy in the axial motion to pass the analyzing plane. In an intermediate range of proton energies, only a fraction of protons can pass, i.e., the ones whose momenta are most aligned with the magnetic field. The transmission function can be analytically calculated in the adiabatic approximation, and the input parameters needed are the electrostatic potential and magnetic field values in both fiducial volume and analyzing plane. As an example, the transmission function for $U=375$\Unit{V} is shown as the dashed dotted line in \fref{fig:ProtonSpectrum}. A detailed field map is only needed to ensure that the field is sufficiently smooth for the adiabatic approximation to be valid.

     Only protons are detected in the $a$SPECT spectrometer, while
     electrons are treated as background. This has the advantage that a precise determination of the electron detection efficiency is not needed, but also the disadvantage that little information about other sources of
     background is directly available. The background subtraction procedure works well for background sources outside the spectrometer. The main problem turned out to be the background generated from trapped particles (electrons or ions) in the Penning-like traps in the center of the spectrometer \cite{Bae08, Sim09}.




The goal for the analysis of the data taken in the recent beam time with $a$SPECT in summer 2013 is to get a first physics result for $a$ with accuracy in the order of a few percent. Gl\"uck \etal have argued \cite{Glu05} that a measurement of $a$ is possible with $a$SPECT with  $\Delta a/a=\SCIe{3}{-3}$. The $a$SPECT spectrometer is planned to become part of the PERC instrument, described in section \ref{sec:BetaAsymmetry}. \cite{Kon12} gives a first discussion of the expected uncertainties.

\subsection{The neutrino-electron correlation {\normalfont \bfseries \itshape a} from electron energy and proton time-of-flight}

\begin{figure}[!ht]%
   \begin{center} \includegraphics{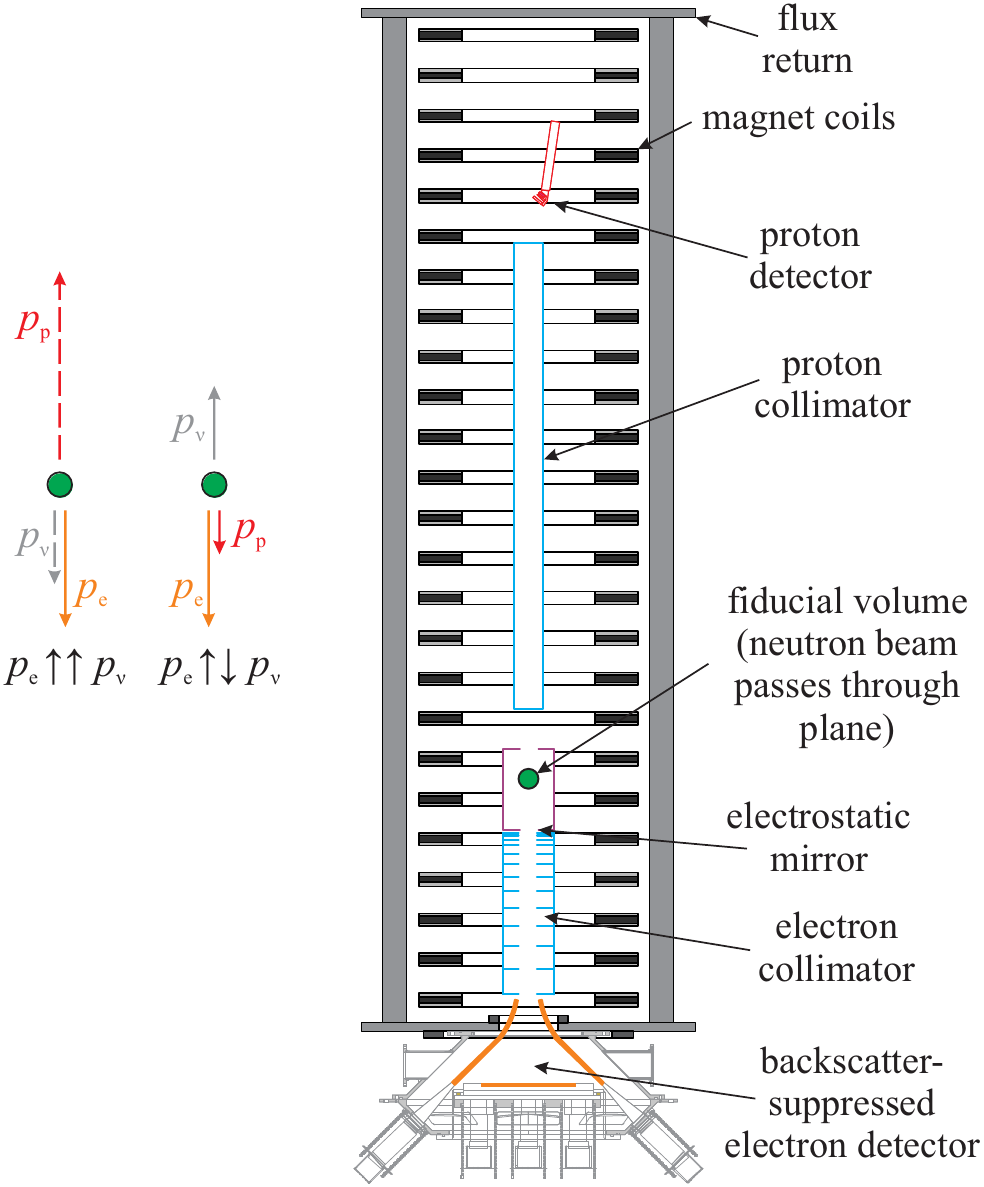} \end{center} 
   \caption{Principle of the aCORN spectrometer. Two momentum diagrams are shown on the left. Both parts of the figure are explained in the text.} 
   \label{fig:aCORNSketch}
\end{figure}

The aCORN collaboration chose a different way to measure the neutrino-electron correlation coefficient $a$ without the detection of the neutrino; their method relies on the three-body kinematics of the decay of the free neutron.  The left part of \fref{fig:aCORNSketch} shows the momenta of decay particles for kinematics in which the electron and neutrino momentum vectors are aligned or anti-aligned. In the aCORN setup, electrons and protons will be detected only if their momenta are close to parallel to the spectrometer axis, and, for the electron, if its momentum points downward to the bottom detector. Since the neutron is decaying essentially at rest, momentum conservation forces the transverse component of the neutrino momentum to be small, too. The neutrino momentum might be mostly parallel (the left momentum diagram) or antiparallel (the right momentum diagram) to the electron momentum; the second possibility is preferred for a negative value of the $a$ coefficient.

\begin{figure}[!ht]%
\begin{center} \includegraphics{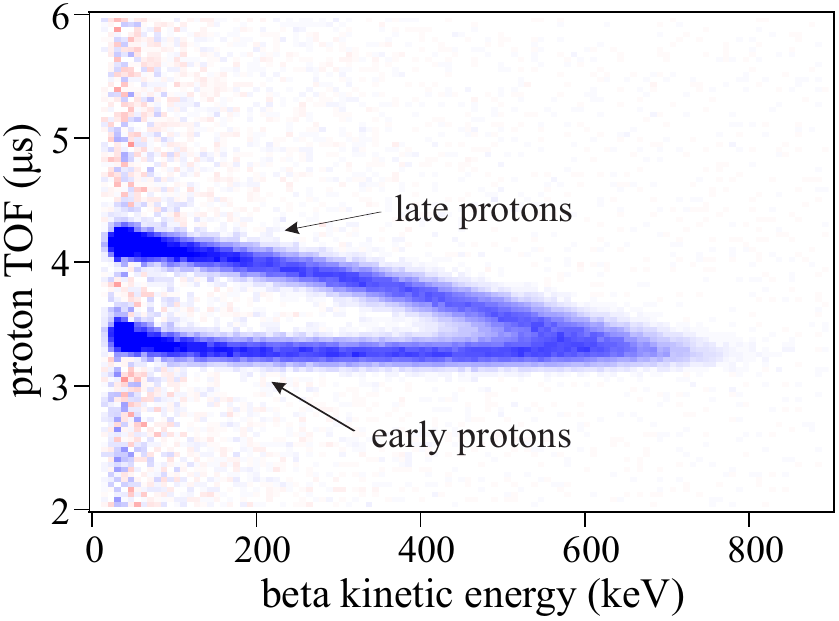} \end{center}
   \caption{Proton TOF spectra, after background subtraction, as a function of electron energy, from \cite{Wiet14p}. The blue bins show a positive count rate, which is higher where the blue is darker. The red bins contain a negative count rate. Most events are in a faster and are slower group. Good separation between the groups is achieved for low and medium electron energy.}  
   \label{fig:aCORNDalitz}
\end{figure}

The idea of the aCORN spectrometer was presented in \cite{Bal94}, and its realization was described in \cite{Wiet05, Wiet09}. The setup is shown in \fref{fig:aCORNSketch}. Neutrons decay in a weak magnetic field of $B\simeq 40$\Unit{mT}. The resulting protons and electrons gyrate around the magnetic field lines with a typical gyration radius of $r_{\rm gyr} \simeq 50$\Unit{cm}. Proton and electron collimators are narrower than $r_{\rm gyr}$, and reject all protons and electrons but the ones whose momenta are close to parallel to the magnetic field. Therefore, the proton's axial movement is either quick and upwards toward the proton detector, or slow and possibly toward the electron detector. In the latter case, an electrostatic mirror voltage reflects it back to the proton detector. The point is that for each electron energy except for the largest ones, there are two groups of protons arriving at the proton detector, one earlier (for initial electron and neutrino momentum parallel) and one later (antiparallel). Strong background reduction comes from the requirement of detecting both the proton and the electron in coincidence. The proton time-of-flight (TOF) spectrum as a function of electron energy is shown in \fref{fig:aCORNDalitz}. The separation of early and late protons disappears for high electron energies, since in this case the neutrino momentum is so small that its direction is no longer relevant. The extraction of $a$ is performed for medium electron energies  (and corresponding speeds $v_{\rm e}=\beta_{\rm e}c$). For each value of electron energy, $a\cdot\beta_{\rm e}$ is given by the asymmetry in the count rates of the events in the fast and slow group, times an apparatus constant that is about $0.7$, and that can be determined by Monte-Carlo simulations to better than 1\% \cite{Wiet14p}. This method is more sensitive to $a$ than the measurement of the shape of the proton spectrum, but only a fraction of the neutron decays in the fiducial volume contribute to the signal.

The aCORN collaboration has finished taking data at the NG-6 beamline at NCNR. Data analysis is ongoing, with the present expectation that it will end in a result for the $a$ coefficient with a statistical uncertainty of about 3\% and a systematic uncertainty less than that \cite{Wiet14p}. The collaboration plans to move to the new, more intense beamline NG-C \cite{Cook09} for another run period that will allow them to reach a final result at about 1\% precision for the $a$ coefficient.

\subsection{The Nab spectrometer}
\label{sec:Nab}

\begin{figure}[!ht]%
   \begin{center} \includegraphics{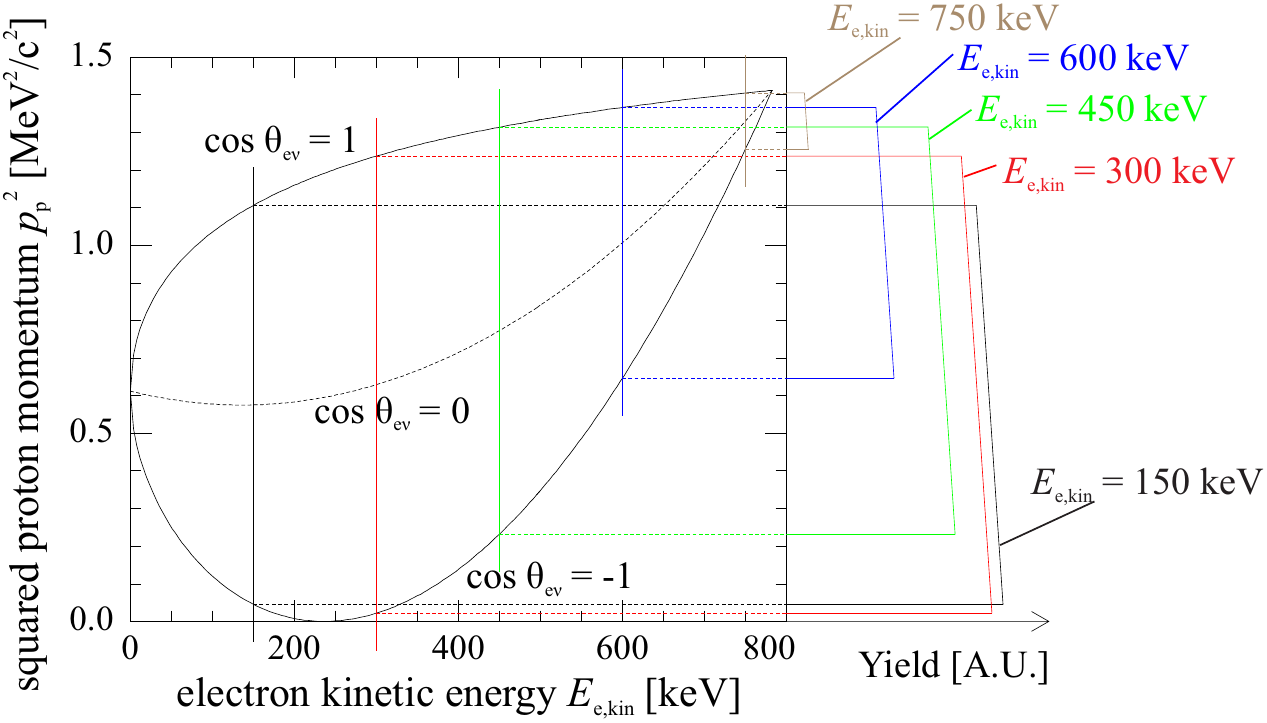} \end{center} 
   \caption{The 2-dimensional plot (left) shows the kinematically allowed region for neutron decays. On the right, we show $p_{\rm p}^2$ distributions for selected electron kinetic energies $E_{\rm e, kin}$. The plot assumes $a=-0.1030(40)$ \cite{PDG12}} 
   \label{fig:NDecay_Dalitz}
\end{figure}

A measurement that uses the information in electron energy and proton TOF in a different way is pursued by the Nab collaboration. The idea of Nab is best explained in the infinite nuclear mass approximation, that is, the masses of the nucleons are taken as infinite, but the neutron-proton mass difference retains its physical value. The proton does get a nonzero recoil momentum, but no recoil energy in neutron decay. Electron momentum $p_{\rm e}$ and, in this approximation, neutrino momentum $p_\nu$ are functions of the relativistic electron energy $E_{\rm e}$. Momentum conservation allows one to compute $p_{\rm p}^2$ as a function of $\cos \theta_{\rm e \nu}$, the cosine of the angle between $\vec p_{\rm e}$ and $\vec p_\nu$:
$p_{\rm p}^2=p_{\rm e}^2+p_\nu^2+2p_{\rm e}p_\nu\cos \theta_{\rm e \nu}$.
For each value of $\Ee$, maximum and minimum values for $p_{\rm p}^2=\KlB{p_{\rm e} \pm p_\nu}^2$ are obtained from $\cos \theta_{\rm e \nu}=\pm 1$. \Fref{fig:NDecay_Dalitz} shows the allowed range of values for $p_{\rm p}^2$ and $E_{\rm e, kin} = \Ee-m_{\rm e}$.
%

Squared proton momentum distributions $P_{\rm p}(p_{\rm p}^2)$ for fixed electron energies $E_{\rm e, kin}$ are shown to the right in \fref{fig:NDecay_Dalitz}. These distributions are given by:
\begin{equation}
P_{\rm p}(p_{\rm p}^2) \propto \left\{
\begin{array}{cl}
\Kla{1+a\frac{p_{\rm e}}{E_{\rm e}}\frac{p_{\rm p}^2-p_{\rm e}^2-p_\nu^2}{2 p_{\rm e} p_\nu}}
& \textrm{if } -1 \le \frac{p_{\rm p}^2-p_{\rm e}^2-p_\nu^2}{2 p_{\rm e} p_\nu} \le 1\\
0 & \textrm{otherwise } \\
\end{array} \right.   \label{eq:pp2Distribution}
\end{equation}
The edges of the $P_{\rm p}(p_{\rm p}^2)$ histograms are sharply defined. Values of $E_{\rm e, kin}$ and $p_{\rm p}^2$ are determined for each
     neutron decay in the Nab experiment, and histograms of squared proton momentum distributions for fixed electron energies are created. The $a$ coefficient is determined from the slope of these histograms.  

\begin{figure}[!ht]%
  \begin{center} \includegraphics{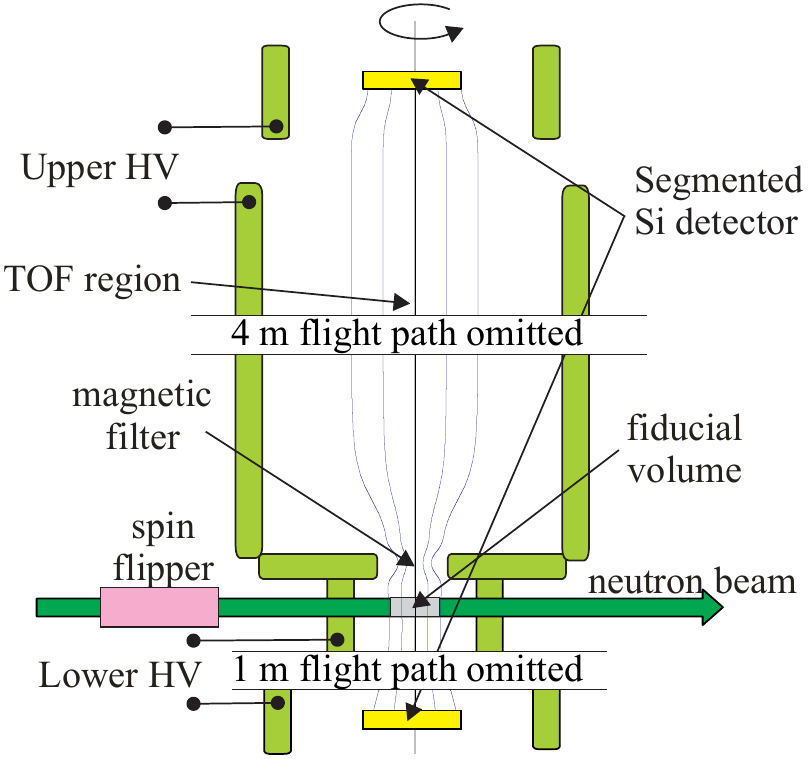} \end{center} 
\caption{Principle of the Nab spectrometer in the vertical orientation. Magnetic field lines (shown in blue) electrodes (light green boxes), and coils (not shown) possess cylindrical symmetry around the vertical axis. The neutron beam is unpolarized for Nab, but can be polarized for later experiments.}
 \label{fig:NabAsymSketch} 
\end{figure}

The original idea of the Nab spectrometer is described in \cite{Bow05,Poc09}. 
The Nab spectrometer was originally planned to have two identical arms at opposite sides of the neutron beam (the symmetric design). Since then, the Nab collaboration has identified a superior strategy; here we report on the final, asymmetric variant of the Nab spectrometer. The asymmetric configuration of the Nab spectrometer is shown conceptually in \fref{fig:NabAsymSketch}.  This version, presently under construction, works as follows:
      Cold neutrons from the FNPB at the SNS pass through the spectrometer. A small percentage of them decay in the fiducial volume, which is in a region of moderate magnetic field, about $1.7$\Unit{T}. Decay protons have to pass through a
field pinch (the filter region) above the fiducial volume to be detected
in the upper detector, the only detector that detects protons.
All electrons, but only upward-going decay protons are accepted. The measured proton time of flight gives an estimate of the squared proton momentum. 

The Si detectors measure the electron energy with keV-level
resolution. Energy losses due to backscattering of
electrons are reduced because the magnetic guide field causes every
electron to be absorbed in one of the two Si detectors, regardless of
which detector is hit first. 

We next turn to the relationship between $t_{\rm p}$, the proton TOF, and $p_{\rm p}$, the proton momentum.
In the adiabatic approximation the proton gyrates around a guiding center which is essentially a magnetic field line. Then, $t_{\rm p}$, the proton TOF, is given by:
\begin{equation}
    t_{\rm p} = \frac{m_{\rm p}}{p_{\rm p}} 
           \int \frac{dz}{\sqrt{1
             -\frac{B(z)}{B_0}\sin^2\theta_0-\frac{e(U(z)-U_0)}{T_0}}} \,,
  \label{eq:fDef_woReflection}
\end{equation}
where $m_{\rm p}$ is the proton mass, $p_{\rm p}$ and $T_0$, are the initial magnitudes of proton momentum
and kinetic energy, while $B_0$ and $U_0$ are the magnetic
field and electric potential, respectively, at the point of decay. The integral is taken along the guiding center. Unfortunately, the angle of the initial proton momentum with the magnetic field, $\theta_0$, is not observable, leading to an imperfect reconstruction of the squared proton momentum $p_{\rm p}^2$ from the inverse squared proton TOF, $1/t_{\rm p}^2$. The influence of the emission angle $\theta_0$ is small in the regions of the spectrometer where the magnetic field is small.
Hence, the most important property of the spectrometer influencing the width of the 1/TOF$^2$ response function, defined as the distribution of $1/t_{\rm p}^2$ values for fixed $p_{\rm p}^2$, is the dispersion in the
time spent between the fiducial volume and the upper end of the filter region.

In the original symmetric configuration \cite{Bow05,Poc09} the beam neutrons decayed in the region with the maximum magnetic field.  The requirements for a large fiducial volume and a
narrow response function were thus in sharp conflict. The relative width of the 1/TOF$^2$ response function response function, 4\%, is determined primarily by
the sharpness of the magnetic field pinch and the the large total flight path length.
The 1/TOF$^2$ response function leads to a small but significant correlation between the value of $a$ and its shape.  Making the
response function narrow decreases the degree to which the proton energy
spectrum is smeared out, and reduces the correlation between the
shape of the response function and the value of $a$.

\begin{figure}[!ht]%
  {\centering \includegraphics{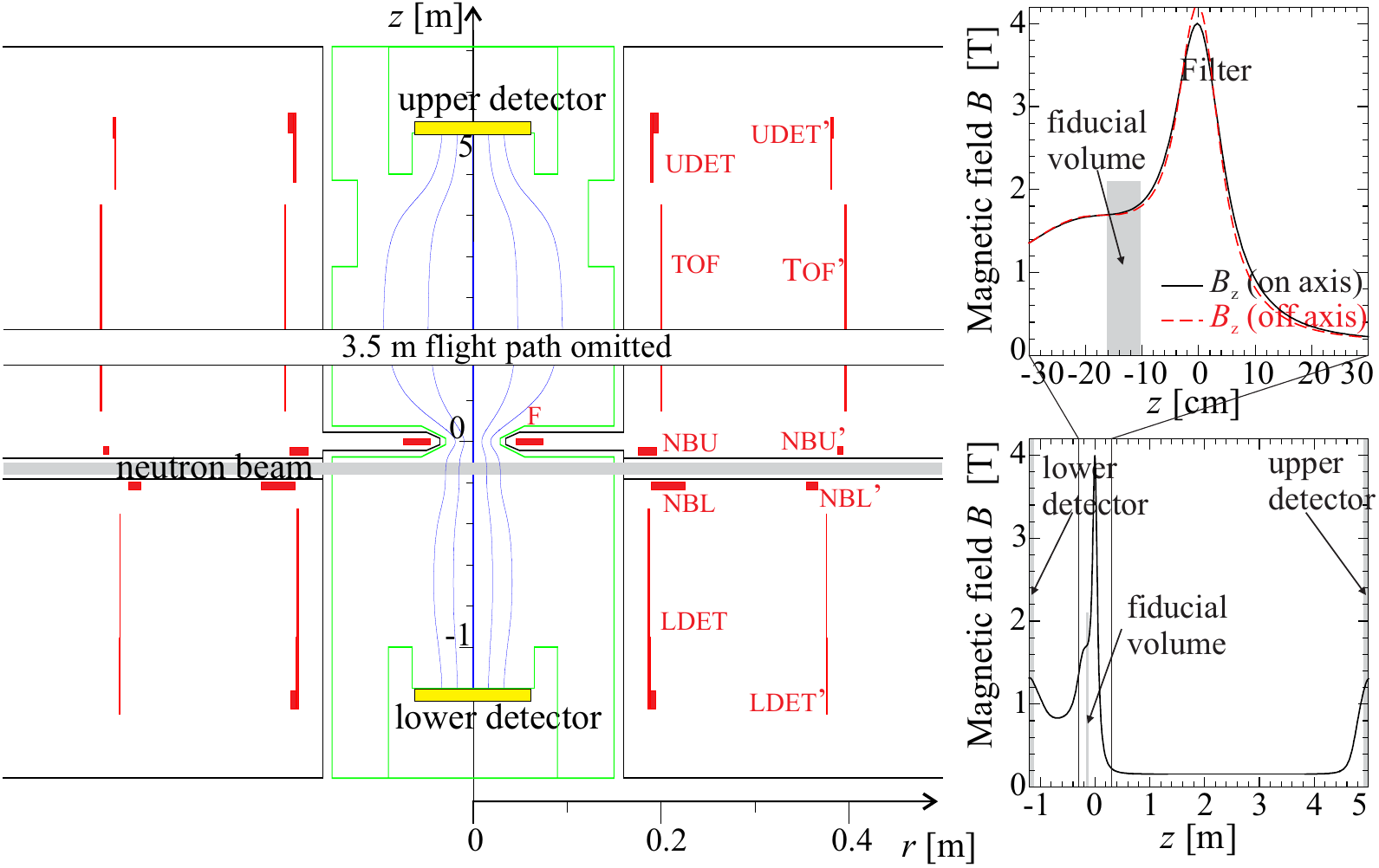}}
    \caption{Detailed view of the Nab spectrometer. Magnetic field lines are shown in blue, electrodes are shown in green, and magnets coils ({\sc UDET}, {\sc TOF}, {\sc F}, {\sc NBU}, {\sc NBL}, and {\sc LDET}) and their counter coils ({\sc UDET$^\prime$}, {\sc TOF$^\prime$}, {\sc NBU}$^\prime$, {\sc NBL$^\prime$}, and {\sc LDET$^\prime$}) are shown as red regions. Much of the flight path is omitted. The diagrams on the right-hand side depict the magnetic field profiles for the full height of the
      spectrometer (bottom), and in more detail for the critical filter
      region (top).}  \label{fig:NabMagnetSystem}
\end{figure}%

The superconducting magnet coil system of the spectrometer is shown in \fref{fig:NabMagnetSystem}.  The coil system has cylindrical symmetry around the vertical symmetry axis. The system is actively shielded to reduce stray magnetic fields: each outer coil has a current orientation opposite to the corresponding inner coil. In addition, the entire apparatus will be enclosed in a passive magnetic
shield made of steel (not shown in the drawing). The lower flight path, not used for the TOF measurement, is relatively short, while the upper flight path, used to measure the proton TOF, is long, leading to a narrow response function.  The source height contribution to the width of the resolution function is minimized because the range of accepted proton directions with respect to the magnetic guide field is significantly narrowed by the magnetic field pinch filter. The neutron beam is kept at a higher magnetic field than the detectors, in order to align the electron momenta to be close to normal to the detector surface when they hit the detector.
In the adiabatic approximation, all protons with $\cos\theta_0$ above a minimum value $\cos \theta_{\rm min }  = \sqrt {1- r_{{\rm B,DV}} }$ pass the filter, where
$r_{\rm B,DV} = 1.7\Unit{T}/4\Unit{T} = 0.425$ is the ratio between the magnetic fields in decay
volume and filter region. This angular cutoff causes only one upward-going proton to be detected for approximately
       eight neutron decays. 

\begin{figure}[!ht]%
\begin{center} \includegraphics{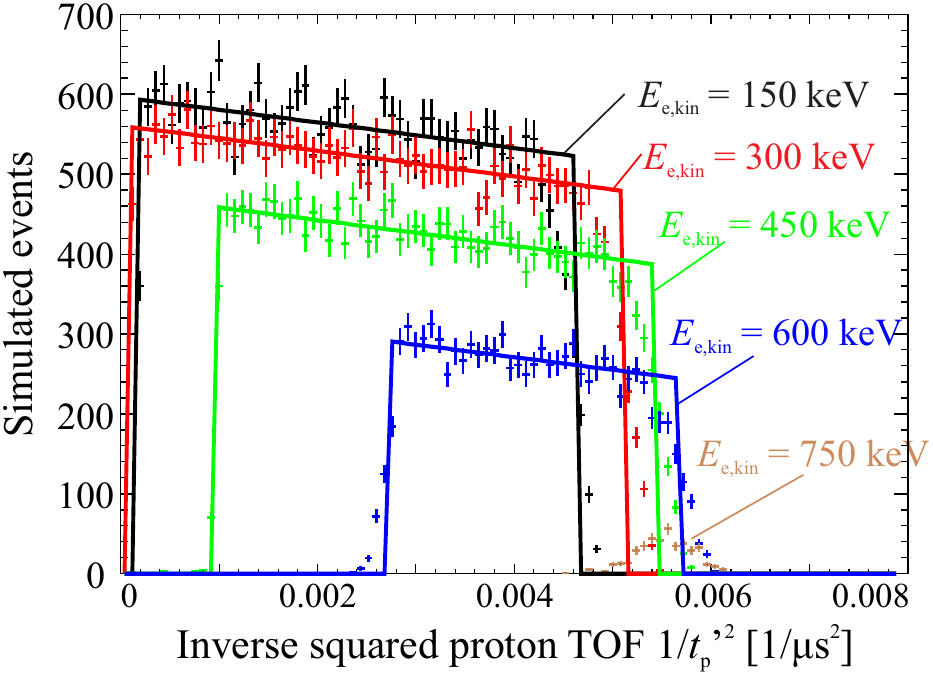} \end{center}
\caption{Proton $1/t_{\rm p}^{\prime 2}$ spectrum for different values of 
       fixed electron kinetic energy $E_{\rm e,kin}$.  The proton TOF has been corrected for the effect of the acceleration region. The solid lines show the expected infinitely sharp 1/TOF$^2$
     response function for an ideal detector. } 
    \label{fig:tp2Spectrum}
\end{figure}

Figure~\ref{fig:tp2Spectrum} shows the proton $1/t_{\rm p}^2$ spectrum
for different electron energies $\Ee$.  The data points are the
results of a realistic GEANT4 simulation that tracks protons through the magnetic fields and the detectors. The data are shown as a histogram in $1/t_{\rm p}^{\prime 2}$, where $t_{\rm p}^{\prime}$ is the simulated $t_{\rm p}$ with a 1-2\% correction that takes into account the electrostatic acceleration at the upper end of the flight path. 
The solid lines show the expectation
for $1/t_{\rm p}^{\prime 2} \propto p_{\rm p}^2$, that is, for a detector 1/TOF$^2$ response function which is infinitely sharp. The $1/t_{\rm p}^{\prime 2}$ spectrum has a
slope proportional to $a$, the neutrino-electron correlation
coefficient, within the range of $1/t_{\rm p}^{\prime 2}$ allowed by the decay
kinematics, and drops sharply to zero outside. The
data points at the edges can be used to determine the shape of the real detector 1/TOF$^2$ response function experimentally. A compromise had to be found in the field design between a very high field curvature in the filter region
(i.e., a narrow detector 1/TOF$^2$ response function), and a
reasonably big fiducial volume, i.e., a filter coil ({\sc F}) in \fref{fig:NabMagnetSystem} that is not too small in diameter.

The
detector has to stop and detect the energy of
up to 750\Unit{keV} electrons as well as 30\Unit{keV} protons.  This
requires a detector thickness of $1.5 - 2$\Unit{mm} Si-equivalent, a
very thin window technology, and a very low energy threshold for
detecting signals down to about 10\Unit{keV}.
The very thin window, or dead layer, must be uniform over a large
area of $\simeq$100\Unit{cm$^2$}. The detector has to be segmented
into 127 elements, in order to keep the individual element small enough to avoid excessive electronic noise, and to allow for rejection of accidental background through the requirement of spatial coincidence between electron and proton hit. The segmentation has been applied on the
back side to keep the irradiated front side homogeneous. The
detector segmentation has to be combined with pulse processing
electronics allowing for real-time signal recording with a resolution
at the level of several ns.  The low energy threshold is required for
good energy resolution, at the level of a few keV for the relevant
energy range of electrons and protons.

Detector prototypes have been procured from Micron Semiconductor
Ltd. \cite{Micron}. The prototypes can be operated at a temperature of
about 100\Unit{K} to reduce electronic noise.  
Charged particles enter the detector through the
junction side, which is uniform apart from a very thin aluminum grid
deposited on it with areal coverage of 0.4\%. Electric charge liberated by the
ionizing particles is collected on the ohmic side, which has 127 individual readout electrodes in the shape of regular hexagons, with area $A_d=70$\,mm$^{2}$. The readout electrodes fill the
      circular area of the detector efficiently, and match the
      image of the fiducial volume well.  Only three hexagons
      meet at a single vertex, thus reducing the maximum number of
      elements involved in a charge-sharing event.  There are no dead spaces between the
      detector elements. 

The performance of prototype Si detectors for low-energy proton detection has been tested, and is described in \cite{Salas14}.

\begin{figure}[!ht]%
\begin{center}  \includegraphics{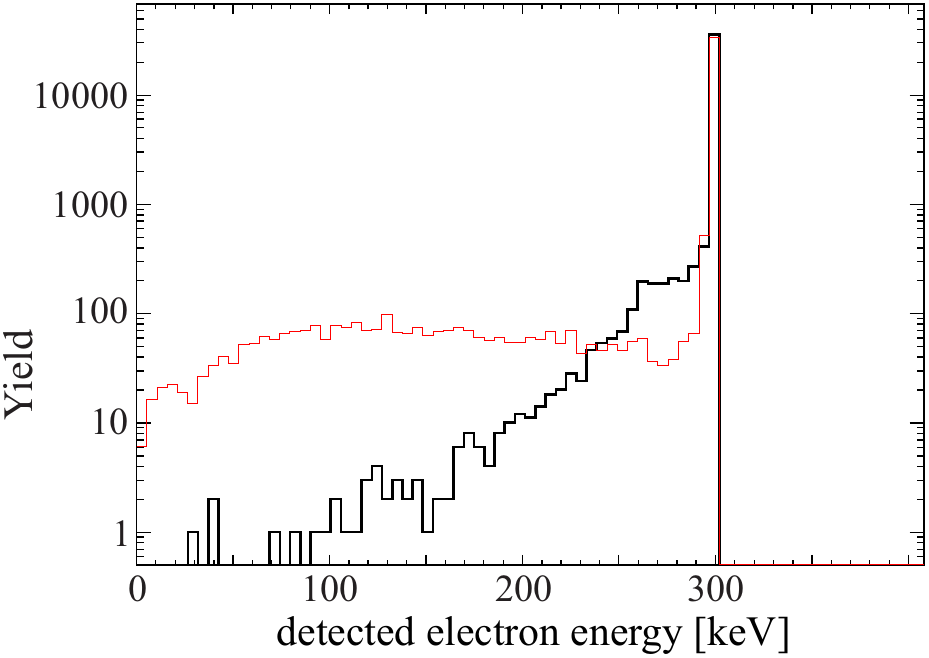} \end{center} 
\caption{Detector response to an incoming electron with a kinetic energy of 300\Unit{keV}, as simulated by GEANT4. The thin red line shows the result if only the lower detector is used. Its long low-energy tail is mainly due to backscattering. The thick black line shows the reconstructed energy in the detector setup as planned in Nab. The algorithm used to reconstruct the energy is explained in the text.}  
    \label{fig:EeResponse}
\end{figure}%

Fitting the data to \eref{eq:pp2Distribution} assumes a perfect determination electron energy $E_{\rm e}$ in the Si detectors. \Fref{fig:EeResponse} shows the simulated electron detector response for monoenergetic incoming electrons. The thin red line shows the response if only the lower detector is used, and backscattering is not suppressed. The thick black line shows the reconstructed energy in a scenario which poses the least demands on the detector readout electronics. Most of the events are contained in a sharp peak at the correct electron energy, reflecting the good energy resolution of a silicon detector. The low-energy tail is mainly caused by electron bremsstrahlung in the detector, with a small additional contribution through energy losses at the dead layer and grid on the detector entrance. Reconstruction of the electron energy requires some care for electrons that bounce multiple times, in order to take into account the different electrostatic potentials on upper and lower detector. Assuming that individual detector hits are not resolved, and that the threshold can be set to 10\Unit{keV} at the upper, and to 40\Unit{keV} at the lower detector, the black line shows reconstructed electron energy using an algorithm that adds the energies in the detectors if they are above threshold, and adds 30\Unit{keV} (due to the high voltage of the upper detector) if the latest detected hit is on the upper detector.

It is planned to suppress proton reflections from the lower detector, e.g., through a $-1$\Unit{kV} voltage at the lower detector. This is necessary, as otherwise too many \cite{Eck78} downward-going protons would bounce off the lower detector undetected, and reach the upper detector within the time window for upwardgoing protons.
Together with the 30\Unit{kV} at the upper detector, and the magnetic field lines connecting the two, the spectrometer constitutes a Penning trap. Experience gained in the commissioning of the $a$SPECT \cite{Sim09} and KATRIN \cite{Beck10} spectrometers indicates different ways to deal with background and HV problems through electrode design, good vacuum, or via the removal of trapped particles by means of electric fields or by mechanical means.

Uncertainties in the determination of $a$: 
The neutrino-electron correlation coefficient $a$ is determined in a
$\chi^2$-fit to the histograms shown in \fref{fig:tp2Spectrum}. 
The fitting parameters are: $a$, $N$, the
number of decays, $b$, the Fierz interference term, and eventually
additional parameters which characterize spectrometer properties.
The knowledge of the spectrometer properties, as determined from auxiliary measurements, is
converted into a prediction of the shape of the the detector 1/TOF$^2$
response function. The fit will be performed twice. First, the full data set is used. In this step, there are two additional fit parameters, the average value of the detector 1/TOF$^2$
response function (parametrized as flight path length $L$), and an
electron energy calibration factor ($E_{\rm cal}$).  In a second step,
the additional parameters are fixed, and the data set is reduced by taking
only the inner 70\% of the data for each given electron energy.  This procedure makes use of the fact that the edges of 1/TOF$^2$ distribution are primarily 
sensitive to the shape of the detector 1/TOF$^2$ response function, and the inner part is mainly sensitive to the neutrino-electron correlation coefficient $a$. The second step reduces the sensitivity of the extracted value of $a$ to imperfections in the reconstruction of the detector function. The decision to use only the inner 70\% of the data sample in the second fit reduces the statistical sensitivity by 65\%, which is offset by considerably
relaxed tolerances for a number of false effects (sources of systematic uncertainty).

\begin{table}[!ht]
  \caption{Statistical uncertainty in the determination of $a$, and the influence of different cuts. $N$ is the number of neutron decays in the fiducial volume for which the proton goes into the upper detector.}  
  \label{tab:StatUncertainty_a}
\begin{center}
\begin{tabular}[c]{ccccc} 
\br
lower $E_{\rm e,kin}$ cutoff: & none & 100\Unit{keV} &
 100\Unit{keV} & 300\Unit{keV} \\ 
upper $t_{\rm p}$ cutoff: & none & none & 40\Unit{$\mu$s} & 40\Unit{$\mu$s} \\
 \mr \\[-12pt]
 $\sigma_a$ & $2.4/\sqrt{N}$ & $2.5/\sqrt{N}$ & $2.5/\sqrt{N}$ & $2.6/\sqrt{N}$ \\
 $\sigma_a$ ($E_{\rm cal}$, $l$ variable) & $2.5/\sqrt{N}$ & $2.6/\sqrt{N}$ & $2.6/\sqrt{N}$ & $2.7/\sqrt{N}$\\
 $\sigma_a$ (inner 70\% of data) & $4.1/\sqrt{N}$ & $4.1/\sqrt{N}$ & $4.1/\sqrt{N}$ & $4.1/\sqrt{N}$ \\
\br
\end{tabular}
\end{center}
\end{table}

The statistical uncertainty is shown in
\tref{tab:StatUncertainty_a} for several possible threshold
values for the electron kinetic energy ($E_{\rm{e,kin,min}}$), and
$t_{\rm{p,max}}$, a high proton TOF cutoff due to accidental
coincidences. Expected settings are $E_{\rm{e,kin,min}} = 100$\Unit{keV} and $t_{\rm{p,max}}=40$\Unit{$\mu$s}.
A total of
$\SCIe{1.6}{9}$ detected protons are required to determine $a$ with a relative
statistical uncertainty of $10^{-3}$. Using the measured beam flux at the SNS \cite{Fom14}, the prediction is that a few months of data taking are needed for a single run, without extra measurements devoted to systematic checks.
Setting $b\equiv 0$ for a fit within the Standard Model, would not significantly improve the uncertainty in $a$. 
 
We turn now to the discussion of the most prominent systematic uncertainties in the determination of $a$ with the Nab spectrometer. A more complete discussion is given in a forthcoming technical publication of the Nab collaboration. The goal of the collaboration is to reduce the total systematic uncertainty to be $\Delta a/a < 10^{-3}$. 
\begin{list}{(\alph{enumi})}{\usecounter{enumi}}
\item \textsl{Magnetic field: }
The magnetic field will be mapped with a precise Hall probe; the relative precision of a magnetic field value taken will be below the $10^{-3}$ level. The most important field specification is the determination of the curvature of the magnetic field in the filter region. The curvature varies by 10\% between the value on axis, and the value for protons at the edge of the fiducial volume. The measurement of the magnetic field curvature is a relative measurement, and the biggest uncertainty is likely the determination of the radial edges of the fiducial volume in the magnetic field map. A shift in the magnetic field curvature of 1\% translates into a shift in $a$ of $\KlB{\Delta a/a} = \SCIe{5}{-4}$.
%
%

%

\item \textsl{Homogeneity of electric potential: }
The fiducial volume, as well as the filter and TOF region, is surrounded by grounded electrodes, providing an electric potential that is close to uniform in these regions. However, the
work function of metals is typically $W \sim
4-5$\Unit{eV} \cite{CRC09}.  For a given metal, it depends on the
crystalline orientation at a level of about $0.3$\Unit{eV}.  This
becomes a problem if different surface materials, orientations, or
just ``dirty'' surfaces are present.  Possible inhomogeneities of the
work function at the electrode surface or surface charges influence
the electrostatic potential distribution. 
%
Implementing an optimized surface coating can reduce the inhomogeneity to below 10\Unit{mV} \cite{Rob06}. The tightest specifications are required for the electrostatic potential inhomogeneities in the decay and filter regions. Here, a difference of 10\Unit{mV} between the two regions would cause a relative shift in $a$ of \SCIe{5}{-4}.

\item \textsl{Neutron beam: Unwanted beam polarization: }
Nab uses an unpolarized neutron beam, and the analysis makes use of the fact that protons are emitted isotropically in neutron decays. Unwanted beam polarization could be caused by polarization of neutrons from the incoming neutron guide, although it is not designed to deliver polarized neutrons. The degree of polarization of the incoming neutrons needs to be studied, and eventually, a spin flipper will be operated in the neutron beam line to minimize this contribution. There is also an expected contribution to the beam polarization due to the transverse and the longitudinal Stern-Gerlach effects. The former leads to a spin-dependent width change of the beam while it enters the spectrometer fields, which in combination with a beam collimator results in polarization. The longitudinal Stern-Gerlach effect accelerates neutrons of one spin state decelerates the others. Therefore, the probabilities for both spin states to have a decay in the fiducial volume are slightly different. The Stern-Gerlach effects produce small and predictable effects. As an additional measure of precaution, it is planned to reverse the direction of the main magnetic field periodically, in order to reverse the sign of the Stern-Gerlach effect.

\item \textsl{Main detector: Electron energy resolution: }
Corrections due to the electron energy response function are small, but significant. The main concern is the low energy tail of the electron energy response function, caused by electrons that create bremsstrahlung in the detector, by electron energy losses in dead layer at the entrance of the detector, and by imperfect energy reconstructions due to energy deposition below threshold in case of backscattered electron events that hit multiple detectors and/or pixels. Several percent of the electrons will have a reconstructed energy which is in this tail. The shape of the tail has to be calibrated to an uncertainty of 1\% in order to reach $\KlB{\Delta a/a}=\SCIe{5}{-4}$.

\end{list}

A cross-check and possible improvement of this analysis will be achieved by a reconstruction of the shape, and not only the average position, of the detector 1/TOF$^2$ response function from the edges of the measured 1/TOF$^2$ distribution.  This
approach introduces more free parameters and uses the previously described analysis method to
provide good starting values for the parameters. The choice of the parameters and their
functional relationship to the detector response function are
discussed in \cite{Nab10}.


\section{Measurement of the Fierz term {\normalfont \bfseries \itshape b} in neutron beta decay}
\label{sec:FierzTerm}

The Fierz interference term vanishes in the Standard Model in lowest order. A non-zero value would show as an additional factor in the electron energy spectrum dependent on the electron energy $\Ee$. Beyond the Standard Model, the Fierz term $b_{\rm F}$ in Fermi decays depends linearly on the scalar coupling. Its non-observation in superallowed beta decays leads to a very stringent limit of $b_{\rm F} = -0.0022(26)$ \cite{Har09}. This limit is set by the fact that for larger values of $b_{\rm F}$, an additional contribution to the lifetime of the nuclei under study would appear, depending on the beta endpoint energy, which has not been observed experimentally. The Fierz term $b_{\rm GT}$ in Gamow Teller decays is less tightly bound \cite{Sev06, Bhat12}. In neutron beta decay, a measurement of the Fierz term $b$ would be a measurement of a linear combination of $b_{\rm F}$ and $b_{\rm GT}$ \cite{Jack57}; given the tight limit on $b_{\rm F}$, this is essentially a measurement of $b_{\rm GT}$ and a search for tensor couplings.
A first limit on $b$ from electron energy spectra in neutron beta decays has been reported with ultracold neutrons recently \cite{Hick13}; indirect limits from the influence of a non-zero $b$ on other neutron decay observables were available earlier \cite{Glu95, Kon10, Pat13}. Neutron decays can compete with radiative pion decays \cite{Byc09} in their sensitivity to tensor couplings if they reach or surpass a precision of $\Delta b \lesssim 10^{-3}$ \cite{Bhat12}.

Measuring the Fierz term $b$ with the Nab spectrometer amounts to a precise determination of the shape of the
electron spectrum and its deviation from the $b=0$ shape predicted by
the Standard Model \eref{eq:CorrCoeffDefinition}. Any such deviation will be far more pronounced for low electron momenta than for high momenta. Accurate understanding of background rates, which usually increases with decreasing
pulse size, is crucial. The coincidence method, i.e., detection of both electrons and proton, will help to reduce the background rates. The main contribution to the background is expected to be due to accidental coincidences; the amount and spectral dependence can be determined with the method of delayed coincidences.
For this measurement, the lower detector will be put on a high negative potential of $-30$\Unit{kV}. The region between the magnetic filter and the upper detector will be kept at $+1$\Unit{kV} in order to reflect all protons created in the fiducial volume with an upward momentum, turning them around into the lower detector, which is thus the only detector sensitive to protons. Given that an overall normalization factor reflecting the number of neutron decays in the fiducial volume has to be fitted to the electron spectrum in addition to $b$, the relevant effect of a non-zero value of $b$ is the distortion of the electron spectrum. Therefore, understanding the energy dependence of electron detection efficiency, especially near threshold, and the detector response function for electron energies, shown in \fref{fig:EeResponse}, are crucial as well.  

The statistical uncertainty in $b$ is given in
\tref{tab:StatUncertainty_b}, which is in agreement with the
calculations of \cite{Glu95}.
\begin{table}[!ht]
  \caption{Statistical uncertainty in the determination of the Fierz Interference term $b$, and the influence of an electron energy cut.} 
  \label{tab:StatUncertainty_b}
  \begin{center}
\begin{tabular}[c]{ccccc}  
\br
lower $E_{\rm e,kin}$ cutoff: & none & 100\Unit{keV} & 200\Unit{keV} & 300\Unit{keV} \\
 \mr \\[-12pt]
 $\sigma_b$ & $7.5/\sqrt{N}$ & $10.1/\sqrt{N}$ & $15.6/\sqrt{N}$ & $26.4/\sqrt{N}$ \\
 $\sigma_b$ ($E_{\rm cal}$ variable) & $7.7/\sqrt{N}$ & $10.3/\sqrt{N}$ & $16.3/\sqrt{N}$ & $27.7/\sqrt{N}$ \\
\br
\end{tabular}
\end{center}
\end{table}
Here, again, $N$ is the number of neutron decays. The statistical uncertainty of $b$ increases
quickly with increasing electron kinetic energy threshold because at higher electron energies a non-zero value of $b$ looks similar to a change of the normalization $N$.  Therefore, the fitting
parameters $N$ and $b$ are strongly correlated. It is planned to run with a threshold of $E_{\rm e,kin,min} = 100$\Unit{keV}. 

The second line of \tref{tab:StatUncertainty_b} lists the
statistical uncertainty in $b$ if the energy calibration is determined
from a fit to the beta spectrum.  The statistical uncertainty is not
significantly worse than that obtained with a fixed energy
calibration.

For an overall sample of $5 \times 10^9$ decays, the statistical
uncertainties $\sigma_b$ calculated on the basis of the above table
are in the range of $\sigma_b \sim 10^{-4}$, far better than the Nab goal
for the overall uncertainty of $\sim 10^{-3}$.  It is clear that the main
challenge in the measurement of $b$ will be in the systematics.

\section{The beta asymmetry {\normalfont \bfseries \itshape A} in neutron beta decay}
\label{sec:BetaAsymmetry}

Presently the most accurate method to determine the ratio of the axial vector and vector coupling constants $\lambda$ is through a measurement of the beta asymmetry $A$, the asymmetry in the angular distribution of the decay electron with respect to the neutron polarization vector. For a neutron beam with polarization $P_{\rm n}$, we have a distribution $w(\Ee,\cos \theta_{\rm e, 0})$ of decay electrons as a function of their energy $\Ee$ (and corresponding speed $v_{\rm e}=\beta_{\rm e}c$) and angle $\theta_{\rm e, 0}$ of its momentum with respect to the polarization vector:
\begin{equation}
w\Kla{\Ee,\cos \theta_{\rm e, 0}} \propto \Kla{1+AP_{\rm n}\beta_{\rm e}\cos\theta_{\rm e,0}} \, .
\end{equation}
This is obtained from \eref{eq:CorrCoeffDefinition}, after integrating over the unobserved parameters and with th assumption that the Fierz term $b=0$.
The beta asymmetry is the correlation coefficient that has been measured the most often in neutron beta decay. A substantial gain in precision has been achieved when experimenters started to design their experiments in a way that the corrections to the raw data are minimized. Previously, the determination of the angular acceptance of the electron detector used \cite{Bopp86,Yero97,Liaud97}, or the neutron beam polarization, led to large corrections that were difficult to determine precisely. The first problem was overcome with a geometry likely introduced by Christensen et al. \cite{Chr72}, with two electron detectors connected by a magnetic field and with the fiducial decay volume in between. In this geometry, provided that the magnetic field is uniform over the decay volume, and that it remains constant or is slowly decreasing towards the detectors, each detector covers a solid angle of $2\pi$ (with corrections to that being very small). The second problem was reduced with the advent of supermirror polarizers \cite{Schaer89}, and later with the technique of crossed supermirrors \cite{Kre05}, that produce close to perfect neutron beam polarization. Opaque spin filters using polarized Helium-3, described in \cite{Zim99}, have turned out to be a reliable tool to characterize the neutron beam spin polarization. The PERKEO II experiment gives the most precise result for the beta asymmetry $A$ \cite{Mund13}, closely followed by the UCNA, described in \cite{Young14}.

\begin{figure}[!ht]
\begin{center} \includegraphics{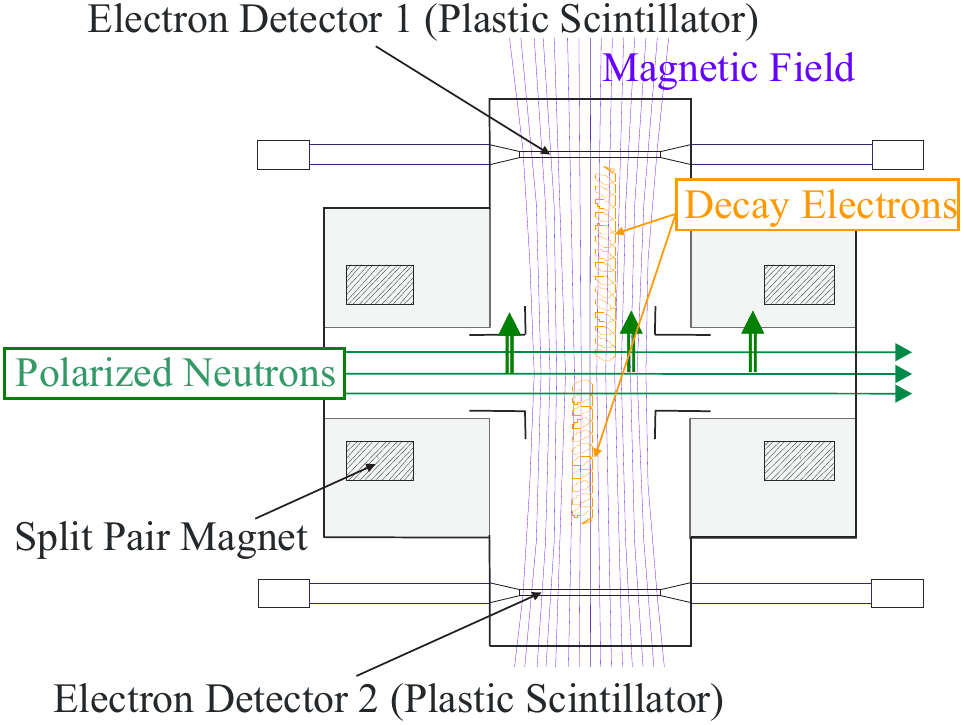} \end{center}
\caption{Sketch of the PERKEO II spectrometer, top view. The main magnet coils have cylindrical symmetry around the magnetic field axis. The operation of the spectrometer is explained in the text.}
\label{fig:PERKEO2_Setup} 
\end{figure}

\begin{figure}[!ht]
\begin{center} \includegraphics{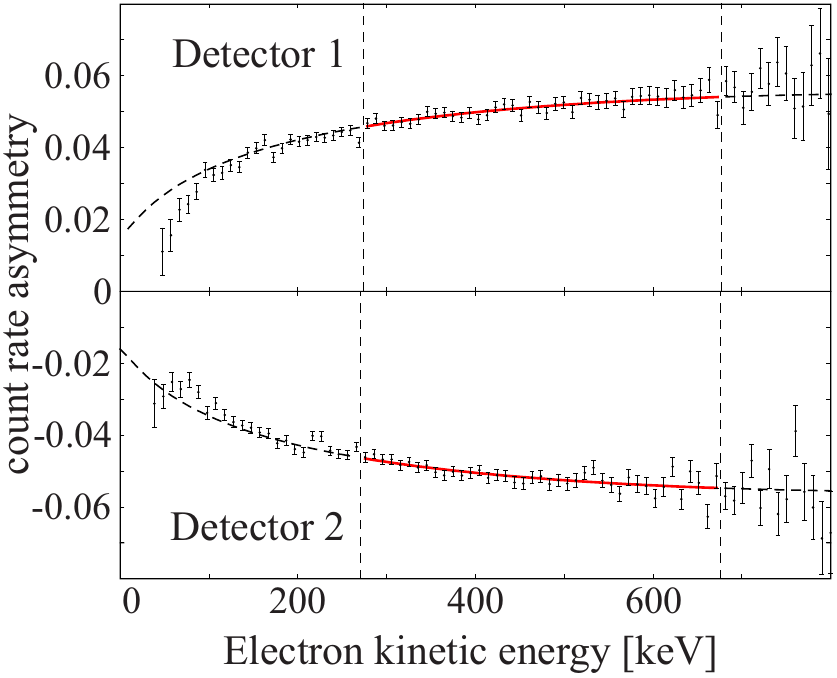} \end{center}
\caption{Count rate asymmetry with respect to the direction of the neutron spin for both detectors, adapted from \cite{Mund13}. The region between the dashed lines is the least sensitive to systematic uncertainties, and is used to extract the result for the beta asymmetry $A$.}
\label{fig:PERKEO2A_Result} 
\end{figure}

The PERKEO II setup is shown in \fref{fig:PERKEO2_Setup}.
In this experiment, a polarized cold neutron beam is directed through the spectrometer. Decay electrons from neutron decays in the fiducial volume in the center of the spectrometer are guided by the magnetic field to one of the two plastic scintillator detectors (read out by photomultiplier tubes)
    at either side of the spectrometer. Each detector covers neutron
    decays into a hemisphere along the neutron spin quantization axis. As a result, each detector measures a count rate asymmetry with respect to the direction of polarization, given by $\pm AP_{\rm n}\beta_{\rm e}/2$, where the factor $\pm 1/2$ comes from the averaging over the electron directions ($\cos \theta_{\rm e, 0}$) in each hemisphere. The results are shown in \fref{fig:PERKEO2A_Result}. 

The largest uncertainties are in the knowledge of the detector response and its spatial dependence, in the estimation of the beam-related background, and in the degree of polarization. The detector response function of the PERKEO II detectors has been calibrated with several conversion electron sources ($^{109}$Cd, $^{113}$Sn, $^{137}$Cs, and $^{207}$Bi). A non-linearity in the gain was found, but could be neglected in the fit region. The position dependence of the detector response has been mapped with some of the calibration sources; in the end an overall gain factor was determined with a fit to the count rate difference between both neutron polarization states, which is taken as background-free. In order to evaluate the measured asymmetry in the count rates, the measured rates are corrected for environmental background by subtracting count rates taken with the neutron beam blocked by a shutter in front of the experiment. This first step of background correction is the largest, but it misses the beam-related background produced by the neutron beam further downstream. An elaborate neutron beam collimation system reduced the amount of beam-related background to give a count rate of less than $1/1700$ of the signal count rate in the fit region, despite the fact that only about one in $10^7$ neutrons decays in the fiducial volume. The size of this background component has been estimated from auxiliary measurements. A decade ago, neutron beam polarization was thought to be the limiting factor in a measurement of the beta asymmetry with cold neutrons. However, crossed supermirrors have made it possible to achieve a very high degree of cold neutron beam polarization, $P_{\rm n} = 99.7(1)$\%. 

\begin{figure}[!ht]
\begin{center} \includegraphics{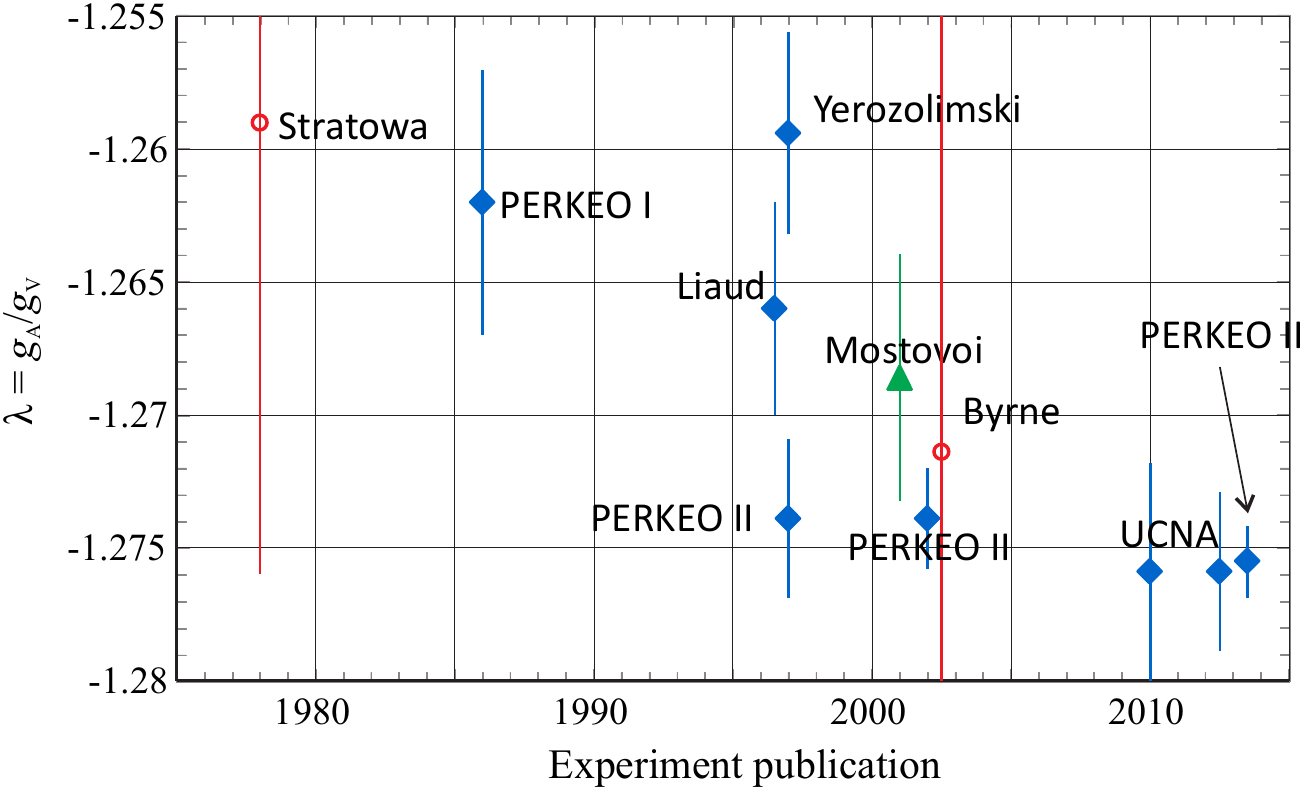} \end{center}
\caption{Compilation of the most precise measurements of $\lambda$.  Data obtained from measurements of the beta asymmetry are shown as blue diamonds, and are taken from UCNA~\cite{UCNA10,UCNA13}, PERKEO II~\cite{Mund13,Abe02,Abe97}, Liaud \etal ~\cite{Liaud97}, Yerozolimsky \etal \cite{Yero97}, and PERKEO I~\cite{Bopp86}. Data obtained from a measurement of beta and neutrino asymmetry is shown as a green triangle, and taken from Mostovo\u{\i} \etal ~\cite{Most01}. Data obtained from measurements of the $a$ coefficient are red circles, and taken from Stratowa \etal~\cite{Str78} and Byrne \etal~\cite{Byr02}.}
\label{fig:LambdaOverview} 
\end{figure}
A value for the ratio of the axial vector and vector coupling constants $\lambda$ can be determined from a beta asymmetry measurement using the relevant equation in \tref{tab:SMValues_CorrCoeff}. At present, selecting the best value involves some judgment, as there is a substantial amount of disagreement between newer and older measurements of the beta asymmetry (see \fref{fig:LambdaOverview}). The experiments described above, $a$SPECT, aCORN, and Nab, that attempt to measure the $a$ coefficient much more precisely than was previously done, are motivated by this discrepancy. The discrepancy also provides a strong incentive to make more precise beta asymmetry measurements.

\begin{figure}[!ht]
\begin{center} \includegraphics{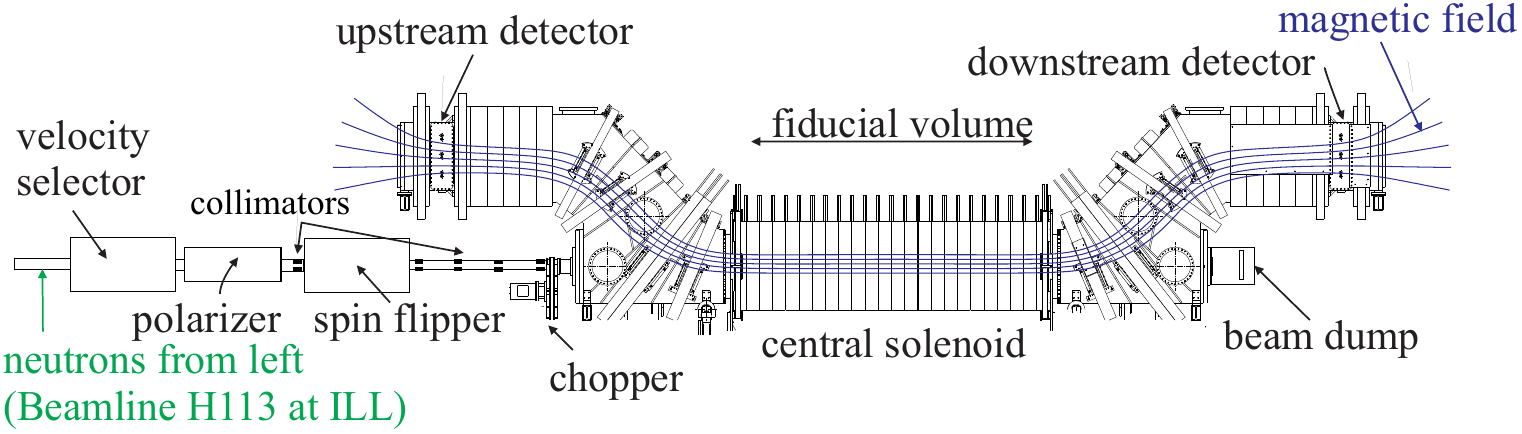} \end{center}
\caption{Sketch of the PERKEO III spectrometer, taken from \cite{Mest11}. Magnetic field lines are the blue lines that connect fiducial volume and detectors. The neutron beam is pulsed due to the operation of velocity selector and chopper in the incoming neutron beam. Data for the analysis is only taken while a neutron beam pulse is contained in the fiducial volume.}
\label{fig:PERKEO3_Setup} 
\end{figure}

The goal of the PERKEO III collaboration is to improve on the precision of PERKEO II. The experiment is described in detail in \cite{Maer09}, and is shown in \fref{fig:PERKEO3_Setup}. 
Like PERKEO II, PERKEO III uses a strong magnetic field traversing a neutron beam and connecting electron detectors at opposite sides. In PERKEO III, the fiducial decay volume has been made much larger by aligning the magnetic field along the neutron beam.  On either side of the fiducial volume, magnetic field lines turn outward to large plastic scintillators which detect decay electrons. The main difference from PERKEO II is the use of a pulsed neutron beam. A small wavelength slice of the neutron beam is selected, as otherwise the velocity spread would lengthen a neutron beam bunch soon after it is made. Data are taken only when the neutron beam bunch is in the fiducial volume of the spectrometer. Besides having a much higher decay rate than in the PERKEO II setup, there are several advantages in the systematics: For one, the relative amount of neutron beam-related background is strongly reduced, as nominally no neutrons hit collimators or the beam stop during data taking. In addition, the neutron beam polarization can be measured more precisely if the wavelength spread of neutrons beam is reduced. Finally, edge effects and the magnetic mirror effect are reduced. The electron detectors are still plastic scintillators read out by photomultipliers. The PERKEO III collaboration has finished data taking, and is finalizing their analysis.

\begin{figure}[!htb]
\begin{center} \includegraphics{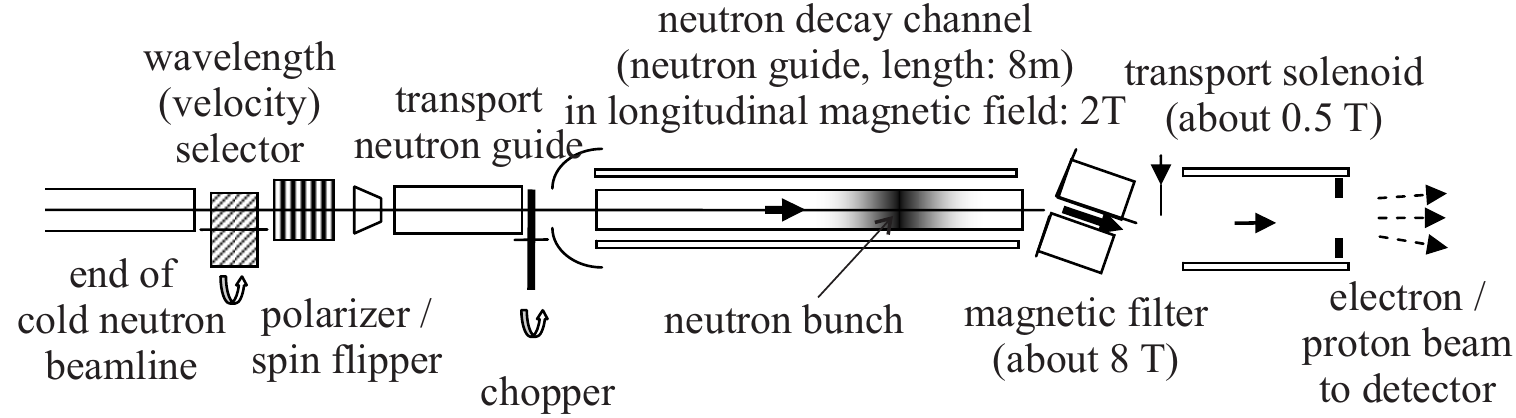} \end{center}
\caption{Sketch of the PERC neutron decay channel setup. Polarizer, spin flipper, wavelength selector, and chopper are optional, and will be used if the neutron beam is desired to be polarized and / or bunched. Adapted from \cite{Dub08}.}
\label{fig:PERCSetup} 
\end{figure}

Looking into the future, construction of the PERC instrument has started at the beam facility MEPHISTO of the Forschungs-Neutronenquelle Heinz Maier-Leibnitz. In contrast to existing neutron decay spectrometers, PERC is a user instrument which delivers, in addition to the neutrons, an intense beam of decay electrons and protons, under well-defined conditions. The setup is shown in \fref{fig:PERCSetup}.  
In PERC, the fiducial decay volume is essentially an 8\Unit{m} long section of neutron guide with a strong magnetic field along the guide. At the end of it, neutrons are directed into a beam stop. Charged decay particles from neutron decays in the 8 m long neutron decay channel follow the magnetic field lines adiabatically, and are guided through a magnetic filter to a detector. This magnetic filter is probably the most important improvement over PERKEO II and III; it was first proposed by Serebrov \etal \cite{SerA05}. One mode to use this instrument is to operate it with a highly polarized bunched neutron beam. Due to the magnetic mirror effect, only decay particles that have their momentum close to parallel to the magnetic field at the decay point can pass the magnetic filter, with the angular cutoff given by the ratio between the magnetic field in the magnetic filter and the magnetic field at the neutron decay point (see appendix). For a wide range of field ratios, the statistical disadvantages of rejecting many decay particles is outweighed by the fact that the decay particles that make it through the filter have the highest asymmetry in their count rate with respect to the neutron spin direction. Coincidence between decay electrons and decay protons is not planned to be used as a method of background rejection, as count rates will be so high that the number of accidentals would be overwhelming. As in PERKEO III,  the amount of beam-related background is suppressed in the data by not recording decay particles when the neutron beam bunch hits the beam stop. A number of systematic effects inherent in the method have been analyzed and found to be sufficiently small, as reported in \cite{Dub08}.
The apparatus needs to be connected to a detector which is specific to the measured observable, e.g., an electron detector for a measurement of the beta asymmetry $A$, or a proton detector for a proton asymmetry. It is possible to use a large plastic scintillator as in the PERKEO experiments. A more innovative proposal has been made by Wang \etal \cite{Wang13} who describe the use of an $R\times B$ spectrometer. Here, magnetic field coils form half of a torus that contains a magnetic field in a semicircle. In this magnetic field, a substantial drift motion along ${\vec R}\times {\vec B}$, where $\vec R$ is the curvature of the magnetic field lines, is superimposed on the particle gyration around a field line as a guiding center. The drift velocity along ${\vec R}\times {\vec B}$ depends on the particle momentum, and therefore the arrival position of a particle on a detector contains information about its momentum. 

Measurements of the degree of polarization in a cold neutron beam well below the present level of 0.1\% appear feasible \cite{Sold14p}. The main limitation is the inhomogeneity in the degree of polarization in different parts of the phase space of the neutron beam. A very high degree of polarization is desirable, as it sets a hard limit on the possible inhomogeneity. 

Looking further in the future, C. Klauser \etal \cite{Klau13} describe an upgrade path, that is, the count rate advantage for the case that the possibilities at MEPHISTO are exhausted and the PERC spectrometer is moved to a planned cold neutron beamline at the European Spallation Source.

\section{The neutrino asymmetry {\normalfont \bfseries \itshape B} in neutron beta decay}

The neutrino asymmetry $B$ describes the asymmetry of the neutrino emission with respect to the neutron spin. While it is not feasible to detect the neutrino with a meaningful efficiency, the value of the neutrino asymmetry can be inferred from the electron and proton momenta. Gl\"uck \etal \cite{Glu95} discuss the two most recently used methods: (1) measurement of the count rate asymmetry for events in which the decay electron and proton are emitted into the same hemisphere relative to the neutron spin -- momentum conservation then restricts the neutrino to the opposite direction, and (2) measurement of the proton asymmetry with respect to the neutron spin. The latter is sensitive to the values of the beta and neutrino asymmetries; however, the beta asymmetry $A$ is known well enough that in practice, a proton asymmetry measurement determines the $B$ coefficient. $B$ is not very sensitive to the value of $\lambda$, and therefore it is measured to search for non-Standard Model physics, e.g., for right-handed currents or scalar and tensor interactions \cite{Glu95, Kon10, Cir13, Hol14}.

\begin{figure}[!ht]
\begin{center} \includegraphics{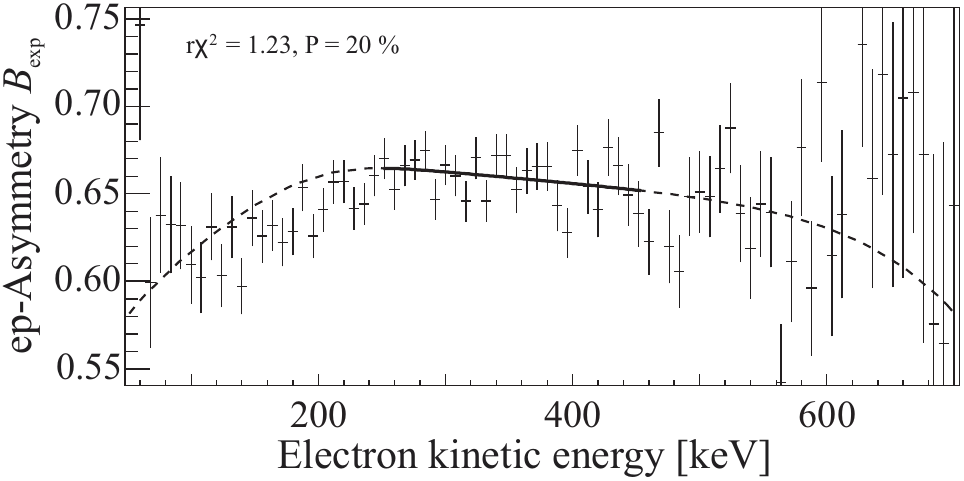} \end{center}
\caption{Asymmetry with respect to the direction of the neutron spin for the count rate of events in which both decay electron and proton are detected by detector 2 in the PERKEO II experiment \cite{Schu07} as a function of the electron kinetic energy. The solid line indicates the fit region that is used to extract the neutrino asymmetry $B$; it is chosen to have the least sensitivity to systematic uncertainties. The dashed lines shows the extension of the fitting curve outside the fit region.}
\label{fig:PERKEO2B_Result} 
\end{figure}

The most precise determinations of the neutrino asymmetry were performed by Serebrov \etal \cite{Ser98}, and by the PERKEO II collaboration \cite{Schu07}. The PERKEO II collaboration modified their instrument that was used for a measurement of the beta asymmetry, as shown in \fref{fig:PERKEO2_Setup} and described above: The main change is the addition of thin carbon foils between the fiducial volume and each of the plastic scintillators; the carbon foils are kept on a high negative potential. The idea is that protons that traverse the foil produce a few secondary electrons; these secondary electrons are detected in the plastic scintillators used for electron detection. Electrons that traverse the thin foil are not very much affected. In this way, the plastic scintillators function as detectors for both electrons and protons. Many sources of background that were a problem for the beta asymmetry measurement were made less significant with a coincidence condition. Electron and proton signals are indistinguishable, creating additional background when one of them is accidentally mistaken for the other. The modified PERKEO II apparatus,  with the carbon foils inserted, was used in a dedicated beam time. The apparatus is most sensitive to the neutrino asymmetry $B$ in a mode counting both electrons and protons in the same detector (see \cite{Glu95}). Results for the asymmetry in the count rate of this class of events with respect to the neutron spin direction from one of the detectors (the better one of the two) are shown in \fref{fig:PERKEO2B_Result}. The fit region is chosen for optimum precision: at higher electron energies the corrections due to magnetic mirror
     effects become large, and at lower energies, high voltage induced
     background and electron detector imperfections generate a large
     uncertainty.

The data recorded with the PERKEO II apparatus just described were additionally used to determine the proton asymmetry with respect to the neutron spin (see \cite{Schu08}). This measurement provides the presently most precise measurement of the proton asymmetry, and has confirmed an earlier result with the same apparatus \cite{Abe05}, as well as the expected dependence of the proton asymmetry on the electron energy. However, the uncertainty in $B$ is considerably higher than with the previous method. The analysis in \cite{Schu08} relies partly on the same data as the one in \cite{Schu07}, and is therefore not an independent determination of the $B$ coefficient.  

Finally we note that Mostovo\u{\i} \etal~\cite{Most01} measured the neutrino and the electron asymmetries in the same device at the same time. In both measurements, the neutron beam polarization $P_{\rm n}$ was not assumed to be known, and therefore, these measurements give a result for $AP_{\rm n}$ and $BP_{\rm n}$. The ratio of these asymmetries is independent of the neutron beam polarization, and it is used to determine $\lambda$ in a different way than others (see \fref{fig:LambdaOverview}).  

\section{Measurement of the beta and the proton asymmetries with the Nab spectrometer}
\label{sec:PolarizedExperiment}

    Use of the Nab spectrometer in experiments with polarized neutrons
    is not planned for the near term, and we would not discuss it here
    were it not for the fact that these experiments use the Nab
    spectrometer with minimal modifications.
    Previously, independent measurements of the electron and proton asymmetry at the FnPB/SNS
    were proposed by the abBA \cite{Wil05, abBA07, Barron10} and PANDA
    \cite{PANDA08} collaborations. Both proposed instruments use
    symmetric magnetic spectrometer designs, with identical flight paths toward
    the two detectors.
    The planned abBA spectrometer had a design with similarities to the
    PERKEO II instrument.  The main differences concerned the type of
    the neutron beam polarizer (a polarized $^3$He spin filter) and
    detectors that allow for proton and electron detection in
    coincidence (the Nab silicon detectors, described in \sref{sec:Nab},
    were originally developed for abBA).
With the Nab spectrometer already under construction, the collaborations joined forces; here we discuss the electron and proton asymmetry measurements that are possible with the Nab spectrometer.  The basic setup shown in \fref{fig:NabAsymSketch} will remain the same, but
     now with a neutron beam polarizing system installed in front of the spectrometer.  For a
measurement of the electron asymmetry it is planned to use a configuration
with the electrostatic voltages set in such a way that all protons are
detected in the lower detector. The coincidence between electrons and
protons from the  same neutron decay is preserved, along with its strong suppression of background-related uncertainties.
The proton asymmetry will be measured with the upper detector serving as the
proton detector, as in the measurement of $a$, the neutrino-electron
correlation.  


For discussion of the uncertainties, we need to consider
an asymmetry of the type 
\begin{equation}
    \textrm{differential decay rate} \propto (1+\alpha \cos\theta_0) \, ,
    \label{eq:GenericAsymmetry}
\end{equation}
where $\theta_0$ is the initial angle of electron
(or proton) momentum relative to the neutron beam polarization (i.e.,
the magnetic field) at the moment of the neutron decay, while $\alpha$
designates the size of the asymmetry, as follows. For the beta asymmetry, the differential decay rate is usually taken as a function of the electron energy, with $\alpha=A\cdot \beta_{\rm e}$.  
\Eref{eq:GenericAsymmetry} defines the proton asymmetry $C$ through $\alpha =2C$ if energy dependencies are not recorded. We note
that it is feasible to measure the proton count rate asymmetry as a
function of the electron energy. In this case $\alpha$ becomes a
function of $A$, $B$ and the electron energy, as given in
\cite{Glu98}, and can be used to make a precise determination of
the neutrino asymmetry $B$. 

The most important sources for uncertainties in a measurement of the electron or proton asymmetry using the Nab spectrometer are as follows:

\begin{itemize}
 \item[1.] {\sl Statistical uncertainty: } 
\Tref{tab:StatUncertainty_A} presents the statistical sensitivity of the beta asymmetry $A$ for different configurations. The likely value for the threshold for the electron kinetic energy, $E_{\rm e,kin,min} = 100$\Unit{keV}, together with the goal for the statistical uncertainty in $\Delta A/A$ of \SCIe{1}{-3}, translates into a demand of $N=\SCIe{1.7}{9}$ neutron decays in the fiducial volume. We note that the neutron beam flux for a polarized beam is lower than the flux of an unpolarized beam due to the polarizer, and depends on the method used to achieve beam polarization. 

 \begin{table}[!ht]
  \caption{Statistical uncertainty in the determination of $A$, and an electron energy threshold. $N$ is the number of neutron decays in the fiducial volume.}  
  \label{tab:StatUncertainty_A}
\begin{center}
\begin{tabular}[c]{ccccc} 
\br
lower $E_{\rm e,kin}$ cutoff: & none & 100\Unit{keV} &
 200\Unit{keV} & 300\Unit{keV} \\
 \mr \\[-12pt]
 $\sigma_A$ & $4.3/\sqrt{N}$ & $4.8/\sqrt{N}$ & $7.8/\sqrt{N}$ & $11.9/\sqrt{N}$ \\
\br
\end{tabular}
\end{center}
\end{table}

If the asymmetry measurements with the Nab spectrometer are performed at the SNS, the use of a $^3$He spin filter is desirable; this is to avoid a large reduction in the neutron beam intensity. If, on the other hand, the Nab spectrometer would move to the new NG-C beamline at NCNR, the higher degree of polarization achievable with the crossed supermirror technique appears to be the better choice. 

\item[2.] {\sl Solid angle: } In a symmetric spectrometer, such as PERKEO II, abBA, or PANDA, the measurement precision relies on the fact that the accepted solid angle of each detector is a hemisphere, and the average angle of electron (proton) momentum with the neutron spin is given as $\overline{\cos \theta_0 } \sim 1/2$ with a correction due to the magnetic mirror effect. 
The asymmetric spectrometer design
   eliminates the unwanted magnetic mirror effect, and replaces it by the requirement to determine the
   solid angle of the upper detector.  The measured electron
   asymmetries in both detectors can be combined in such a way that the angle cutoff of each detector drops out in leading order, or in a different way that allows us to extract the solid
   angle of the upper detector \textsl{in situ} from the beta asymmetries in both detectors, e.g., for subsequent use in a 
   proton asymmetry measurement. The cutoff angle for each detector depends on the magnetic field at the position of  neutron decay. Nevertheless, we show in the appendix \eref{eq:Aexp_asym_korr} that the magnetic field inhomogeneities can be neglected for the Nab setup.  Therefore, no high precision magnetic field measurements are needed, and the systematic uncertainty due to the
   solid angle (as well as the the uncertainty due to the imperfect knowledge of the neutron beam position) is small.
   We note that the usual arrangement with two identical detectors
   used for electron and proton asymmetry measurements is replaced by
   two different detectors, but with a difference which is precisely
   understood.

\item[3.] {\sl Neutron beam polarization: } 
A common systematic to all experiments that measure an asymmetry with respect to the spin direction is the measurement of the beam polarization $P_{\rm n}$. We have argued above in the description of the PERC spectrometer that
   the uncertainty in the beam polarization will be below 0.1\% for a
   cold neutron beam polarized with supermirrors, the level that was
   already achieved by the PERKEO II collaboration.
The same argument is made in \cite{PANDA08} for a $^3$He spin filter. We note that if a $^3$He spin filter is used, the neutron beam polarization depends on the neutron wavelength in a known way. For an experiment at a spallation source, this translates into a time-dependence of the degree of polarization which in itself can provide an in-situ measurement, or at least an in-situ monitor, of the neutron beam polarization \cite{Pen05}.

\item[4.] {\sl Electron energy calibration: } It is planned to use the same detector as in the unpolarized experiments described previously. Its simulated electron energy response has been shown and discussed in \fref{fig:EeResponse}. Its width is substantially smaller than with a plastic scintillator detector.
To determine the detector response function, and to establish the linearity
of the energy-channel-relationship, a set of radioactive calibration sources will be used. The sources are backed by very thin (e.g., 10\Unit{$\mu$g/cm$^{2}$}) carbon
foils, and are movable within the fiducial volume so as to reach every point in the detector.
Six possible candidates for such
calibration sources have been used in ref. \cite{Abe02}. For the electron asymmetry, an uncertainty of 0.1\% in
the slope of the energy channel relation would cause an uncertainty of $\Delta A/A = \SCIe{2}{-4}$ for a reasonable choice of the fitting region.

\item[5.] {\sl Electric field homogeneity: } The proton asymmetry is
  very sensitive to electric potential inhomogeneities; the beta asymmetry is not, as electron energies are much higher. Demands on the absence of shallow electric field minima in symmetric spectrometers like PERKEO II, abBA, or PANDA, are tight.
%
  However, the sensitivity to electric potential traps is practically absent in a
  proton asymmetry measurement using the asymmetric Nab spectrometer.  A trapped electron or proton needs to come from a decay event in this trap. If its kinetic energy is above a low-energy cutoff, it is trapped only because its momentum is close to perpendicular to the magnetic field. Such electrons or protons would not make it through the magnetic filter even in the absence of a trap. The leading contribution to the list of uncertainties is an (unwanted)  electric potential difference between the filter region and the
  fiducial volume.  A filter--fiducial volume potential difference as large as
  500\Unit{mV} changes the proton asymmetry by less than 0.04\% for
  $r_{\rm B,DV} \sim 0.5$.

\end{itemize}

The asymmetric Nab spectrometer design allows for a beta and proton asymmetry measurement at an uncertainty level of $10^{-3}$ or better.

\section{Summary}

The experiments at the Spallation Neutron Source and the planned measurements with the PERC instrument promise a series of precision measurements of correlation coefficients in neutron beta decay at the level of 0.1\% or better in the near future.  These measurements of beta asymmetry and neutrino--electron correlation coefficient, along with a corresponding improvement in the knowledge of the neutron lifetime, will make possible a test of the unitarity of the CKM matrix with an accuracy comparable to the one using superallowed beta decays, but without the nuclear uncertainties. In addition, the search for scalar and tensor interactions in neutron decays, which contribute mainly to the Fierz term and the neutrino asymmetry, will explore new physics with a sensitivity that is at or beyond the level of searches at the LHC. The intention of this review has been to present the experiments that are planned to achieve these goals.

\ack

We thank the members of the Nab collaboration, and H. Abele, B. M\"arkisch, J. Nico, T. Soldner, and F. Wietfeldt, for help in the preparation of the manuscript. We gratefully acknowledge support from the Office of Nuclear Physics in the Office of Science of the Department of Energy, and the National Science Foundation, through NSF grants PHY-0653356, -0855610, -0970013, -1205833 and -1307328.

\appendix
\setcounter{section}{1}

\section*{Appendix: Sensitivity of count rate asymmetry to inhomogeneities in the magnetic field}

\def \nU{_{\rm 0,U}}
\def \nD{_{\rm 0,D}}
\def\Rave{{\overline{r_{\rm B,DV}}}}
\def\Dr{{\delta r_{\rm B,DV}}}
\def\DDr{{\varSB(r_{\rm B,DV})}}

%

The measured asymmetry for a decay rate as in
\eref{eq:GenericAsymmetry} is
\begin{equation}
  \alpha_{{\rm{exp}}}
      = \frac{{N^\uparrow - N^\downarrow}} {{N^\uparrow+ N^\downarrow}}
          = \alpha\,\overline{\cos \theta_0 }  \, .
  \label{eq:Aexp_sym}
\end{equation}
We denote the count rates in the upper detector as $N^\uparrow$ and
$N^\downarrow$ for the different neutron spin states.  In the
adiabatic approximation, decay particles are detected in the upper
detector if their emission angle fulfills the condition $\cos\theta_0
> \cos\theta_{\rm min} = \sqrt{1 - r_{\rm B,DV} }$, where $r_{\rm B,DV}\sim 0.425$ is the ratio between the magnetic fields in decay
volume and filter region. Given that
$\cos\theta_0$ is uniformly distributed, the average emission
angle, for particle that go to the upper detector, is:
\begin{equation}
  \overline {\cos \theta\nU }
        = \frac{1}{2}\left( {1
             + \sqrt {1 - r_{{\rm B,DV}} } } \right)\,.
\end{equation}
In the same way, we determine the average emission angle for particles that go to the lower detector to be
\begin{equation}
  \overline {\cos \theta\nD }
        = -\frac{1}{2}\left( {1
             - \sqrt {1 - r_{{\rm B,DV}} } } \right)\,.
\end{equation}

If both detectors are used to evaluate asymmetries, one can choose a combination independent of $r_{\rm B,DV}$. We add the subscripts ``U'' and ``D'' for count rates and asymmetries in upper
and lower detector, and make use of the fact that the two detectors
see opposite solid angles.  
\begin{equation}
  \alpha_{{\rm{exp,U}}}  - \alpha_{{\rm{exp,D}}}
     = \frac{{N_{\rm{U}}^\uparrow  - N_{\rm{U}}^ \downarrow}}
       {{N_{\rm{U}}^ \uparrow + N_{\rm{U}}^ \downarrow}} 
        - \frac{{N_{\rm{D}}^\uparrow - N_{\rm{D}}^\downarrow}}
               {{N_{\rm{D}}^ \uparrow   + N_{\rm{D}}^\downarrow  }}
     \equiv \alpha \,.
\label{eq:Aexp_asym}
\end{equation}
The sensitivity of this expression to $\alpha$ is about the same as
\eref{eq:Aexp_sym}.

Furthermore, one can determine the average emission angle for both detectors through
\begin{eqnarray}
   \overline {\cos \theta\nU }
     &= \left( {1 - \frac{{\alpha_{{\rm{exp,D}}} }}{{\alpha_{{\rm{exp,U}}} }}}
            \right)^{ - 1}     \label{eq:ThetaDetermination} \\
   \overline {\cos \theta\nD } &= \overline {\cos \theta\nU }-1 \,.
\end{eqnarray}
The determination of the average emission angles is best done with electrons, as $\alpha_{\rm{exp,D}}$ for protons is very sensitive to systematic uncertainties (traps due to a electrostatic potential minima).

Equations \eref{eq:Aexp_asym} and \eref{eq:ThetaDetermination} are strictly valid only if the magnetic field
ratio $r_{\rm B,DV}$ does not depend on the position of the neutron decay. We next turn to the error associated with this simplification: 
%
$n({\vec x})$ is the normalized density of neutron decays, and $\alpha$
is the asymmetry, which may be the electron or the proton
asymmetry, corrected for the kinematic factors and degree of
polarization.  We note that now $r_{\rm B,DV}({\vec x})$ depends on the
position. We write $r_{\rm B,DV}({\vec x})=\Rave+\Dr({\vec x})$ with $\overline{\Dr({\vec x})}=0$. The averaging is understood to be performed over the fiducial volume
weighted by the density of neutron decays.  The count rate $N_{\rm{U}}^ \uparrow$ in the upper detector for neutron-spin pointing upwards is given as:
\begin{equation}
   N_{\rm{U}}^ \uparrow
       = \frac{{N_0 }}{2}\int {d^3x \cdot n({\vec x})}  \cdot
       \int\limits_{\sqrt {1 - r_{{\rm{B,DV}}({\vec x})} } }^1 \Kla{ {1 +
       \alpha \cos \theta_0 }\, d\cos \theta_0 } \,.
\end{equation}
Carrying out the integration, we obtain
\begin{eqnarray}
    N_{\rm{U}}^ \uparrow  & = \frac{{N_0 }}{2}\Klb{ \Kla{1 -
    \overline {\sqrt {1 - r_{{\rm{B,DV}}}({\vec x}) } } } + \frac{\alpha
    }{2}\overline {1 - \Kla{1 - r_{\rm{B,DV}}({\vec x})} } }  \nonumber \\ 
      &=  \frac{{N_0 }}{2}\Klb{ 1 - \sqrt{1-\Rave}\overline{\sqrt{1-\frac{\Dr}{1-\Rave}}}+ \frac{\alpha }{2}\Rave }  \\
& \simeq \frac{{N_0 }}{2}\Klb{ 1 - \sqrt{1-\Rave}\Kla{1-\frac{\DDr}{8\Kla{1-\Rave}^2}}+ \frac{\alpha }{2}\Rave }   \,.
\end{eqnarray}
Analogously, we find for the count rate in the spin-down state of the neutron
\begin{eqnarray}
    N_{\rm{U}}^ \downarrow  & = \frac{{N_0 }}{2}\Klb{ \Kla{1 -
    \overline {\sqrt {1 - r_{{\rm{B,DV}}}({\vec x}) } } } - \frac{\alpha
    }{2}\overline {1 - \Kla{1 - r_{\rm{B,DV}}({\vec x})} } }  \nonumber \\ 
& \simeq \frac{{N_0 }}{2}\Klb{ 1 - \sqrt{1-\Rave}\Kla{1-\frac{\DDr}{8\Kla{1-\Rave}^2}}- \frac{\alpha }{2}\Rave }   \,.
\end{eqnarray}
For the asymmetry in the upper detector $\alpha_{\rm U}$, we expect:
\begin{eqnarray}
  \alpha_{{\rm{exp,U}}}  & = \frac{{N_{\rm{U}}^ \uparrow   - N_{\rm{U}}^
  \downarrow  }}{{N_{\rm{U}}^ \uparrow   + N_{\rm{U}}^ \downarrow  }}
  = \frac{\alpha  \cdot \Rave }{2\Kla{1 -
  \overline {\sqrt {1 - r_{{\rm{B,DV}}} } } } } \nonumber \\ 
  & \simeq \frac {\alpha  \cdot \Rave }{2\Kla{1 -\sqrt{1-\Rave }}}
    \Kla{1-\frac{\DDr\Kla{1-\Rave}^{-3/2}}{8\Kla{1 -\sqrt{1-\Rave }}}}  \nonumber \\ 
  & \simeq \frac {\alpha  \cdot \Kla{1 +\sqrt{1-\Rave }}}{2}
    \Kla{1-\frac{\DDr\Kla{1-\Rave}^{-3/2}}{8\Kla{1 -\sqrt{1-\Rave }}}} \, .
\end{eqnarray}
For the lower detector, we find
\begin{eqnarray}
  N_{\rm{D}}^ \downarrow   & = \frac{{N_0 }}{2}\left[ {\left( {1 +
  \overline {\sqrt {1 - r_{{\rm{B,DV}}} } } } \right) + \frac{\alpha
  }{2}\overline {r_{{\rm{B,DV}}} } } \right] \,, \textrm{and} \\ 
  N_{\rm{D}}^ \uparrow   & = \frac{{N_0 }}{2}\left[ {\left( {1 +
  \overline {\sqrt {1 - r_{{\rm{B,DV}}} } } } \right) - \frac{\alpha
  }{2}\overline {r_{{\rm{B,DV}}} } } \right] \,.
\end{eqnarray}
Finally, for the asymmetry $\alpha_{\rm D}$, we obtain
\begin{eqnarray}
  \alpha_{{\rm{exp,D}}}  & = \frac{{N_{\rm{D}}^ \uparrow   - N_{\rm{D}}^
  \downarrow  }}{{N_{\rm{D}}^ \uparrow   + N_{\rm{D}}^ \downarrow  }}
  = -\frac{\alpha  \cdot \Rave }{2\Kla{1 -
  \overline {\sqrt {1 + r_{{\rm{B,DV}}} } } } } \nonumber \\ 
  & \simeq -\frac {\alpha  \cdot \Rave }{2\Kla{1 +\sqrt{1-\Rave }}}
    \Kla{1+\frac{\DDr\Kla{1-\Rave}^{-3/2}}{8\Kla{1 +\sqrt{1-\Rave }}}}  \nonumber \\ 
  & \simeq -\frac {\alpha  \cdot \Kla{1 -\sqrt{1-\Rave }} }{2}
    \Kla{1+\frac{\DDr\Kla{1-\Rave}^{-3/2}}{8\Kla{1 +\sqrt{1-\Rave }}}} \, .
\end{eqnarray}
These asymmetries have to be subtracted, as shown in
\eref{eq:Aexp_asym}, yielding
\begin{eqnarray}
   \alpha_{{\rm{exp,U}}}  - \alpha_{{\rm{exp,D}}} &
        = \frac{{\alpha \cdot \overline {r_{{\rm{B,DV}}} } }}
            {{2\left( {1 - \overline{\sqrt {1 -
              r_{{\rm{B,DV}}}}}}\right)}} + \frac{{\alpha  \cdot
          \overline {r_{{\rm{B,DV}}} } }}{{2\left( {1 + \overline {\sqrt {1 -
           r_{{\rm{B,DV}}} } } } \right)}} \nonumber \\   
   & = \frac{{\alpha  \cdot \overline{r_{{\rm{B,DV}}} } }}
              {{1 - \left(\overline{\sqrt{1 -
                           r_{{\rm{B,DV}}}}}\right)^{\!\!2}}} \nonumber \\   
   & = \frac{\alpha  \cdot \Rave }{1 -\Kla{1-\Rave}
   \overline{\sqrt{1-\frac{\Dr}{1-\Rave}}}^{\,2} } \nonumber \\   
   & \simeq \alpha \Kla{1+\frac{\DDr}{4\Rave\Kla{1-\Rave}}} \,.
\label{eq:Aexp_asym_korr}
\end{eqnarray}
The equation gives a correction to \eref{eq:Aexp_asym}
for an inhomogeneous field, which is well below $10^{-3}$ for the geometry used in the Nab spectrometer.

Finally, we compute the average emission angle $\overline {\cos \theta\nU }$ as follows:
\begin{eqnarray}
\overline {\cos \theta\nU } &= \Klb{1-\frac{\alpha_{{\rm{exp,D}}}}{\alpha_{{\rm{exp,U}}}}}^{-1}  \nonumber \\
   & = \Klb{1+\frac{\alpha \cdot \Rave}{2\Kla{1-\overline{\sqrt {1 + r_{\rm{B,DV}} } } } }
       \Kla{\frac{\alpha \cdot \Rave}{2\Kla{1-\overline{\sqrt {1 - r_{\rm{B,DV}} }}}}}^{-1} }^{-1} \nonumber \\
   &= \frac{1+\overline{\sqrt {1 - r_{\rm{B,DV}} }}}{2} \nonumber \\
   &\simeq \frac{1}{2}\Kla{1+\sqrt{1-\Rave}}-\frac{\DDr\Kla{1-\Rave}^{-3/2}}{16} \, .    
\end{eqnarray}
Again, for the geometry of the Nab spectrometer, the second term constitutes a correction well below $10^{-3}$, which can be neglected.



\begin{thebibliography}{00}
\bibitem{Hol14}
Holstein B, this focus issue of \jpg (citation will be filled in by editor)
\bibitem{Young14}
Young A, this focus issue of \jpg (citation will be filled in by editor)
\bibitem{Sev14}
Severijns N, this focus issue of \jpg (citation will be filled in by editor)
\end{thebibliography}
\end{document}